\documentclass[pre,twocolumn]{revtex4}
\pdfoutput=1
\usepackage{amsfonts}
\usepackage{amsmath}
\usepackage{amssymb}
\usepackage{graphicx}

\begin{document}
\title{Fluid flow enhances the effectiveness of toxin export by aquatic microorganisms: a first-passage perspective on microvilli and the concentration boundary layer}
\author{Nicholas A. Licata}
\email[Corresponding author address: ]{licata@umich.edu}
\affiliation{Department of Natural Sciences, University of Michigan-Dearborn, Dearborn, Michigan 48128}
\author{Aaron Clark}
\affiliation{Department of Natural Sciences, University of Michigan-Dearborn, Dearborn, Michigan 48128}
\begin{abstract}
A central challenge for organisms during development is determining a means to efficiently export toxic molecules from inside the developing embryo.  For aquatic microorganisms, the strategies employed should be robust with respect to the variable ocean environment and limit the chances that exported toxins are reabsorbed.  As a result, the problem of toxin export is closely related to the physics of mass transport in a fluid.  In this paper we consider a model first-passage problem for the uptake of exported toxins by a spherical embryo.  By considering how macroscale fluid turbulence manifests itself on the microscale of the embryo, we determine that fluid flow enhances the effectiveness of toxin export as compared to the case of diffusion-limited transport.  In the regime of large P\'eclet number, a perturbative solution of the advection-diffusion equation reveals that a concentration boundary layer forms at the surface of the embryo.  The model results suggest a functional role for cell surface roughness in the export process, with the thickness of the concentration boundary layer setting the length scale for cell membrane protrusions known as microvilli.  We highlight connections between the model results and experiments on the development of sea urchin embryos.
\end{abstract}

\maketitle


\section{Introduction}
Aquatic organisms face a variety of challenges in the course of development.  Central challenges related to their growth and development are the acquisition of nutrients from the surrounding fluid and the disposal of waste products or other toxic materials to the extracellular environment.  As a result, acquatic organisms have evolved a diverse set of strategies to search for, acquire and dispose of small molecules.  Successful strategies reflect fundamental constraints imposed by the physical laws which govern the transport and motion of small particles in a fluid.  This line of physical reasoning has shaped our understanding of a variety of problems in biology, from bacterial chemotaxis \cite{berg1977physics,berg1993random} to the origin of multicellularity in algae \cite{goldstein2011evolution,short2006flows}.    

The present paper highlights a connection between the physics of mass transport in a fluid flow and the problem of removing toxic molecules or other waste products from a developing embryo.  The major question addressed can be stated quite simply.  A spherical embryo has identified a toxic molecule for export to the extracellular fluid.  Once exported the molecule will be subject to diffusion and advection in the surrounding fluid.  How far away from the body of the embryo should the molecule be released, so as to reduce the chances that the toxin encounters the embryo surface and is reabsorbed?  The view advocated in the present paper is that the physics underlying this transport problem provides an answer that may shed light on understanding the functional role of cell surface roughness in embryonic development.  Later we will argue that the length of cell surface protrusions known as microvilli (the surface roughness elements implicated in the toxin export process) may be set in part by the thickness of the concentration boundary layer for the advection-diffusion problem.  

Recent experiments on sea urchin development have highlighted the important role that cell surface roughness plays in toxin export.  Early in sea urchin development, microvilli lengthen, and there is a coincident localization of transport receptors to the tips of microvilli \cite{whalen2012actin,gokirmak2012localization}.  These transport receptors act to export toxic molecules from the interior of the cell to the extracellular fluid \cite{lange2001microvillar,lange2011fundamental}.  This suggests that the localization of transport receptors to the tips of microvilli may serve a functional role in the export process.  Releasing the toxic molecules at a distance $h$ (the microvilli length) from the cell membrane surface may reduce the chances that exported toxins are subsequently reabsorbed by the cell.  

In this paper we investigate the efficacy of the tip localization strategy by considering a model first-passage problem \cite{redner2001guide} for the uptake of exported toxins by a spherical embryo.  In Section II we consider the regime of diffusion-limited transport.  We demonstrate that tip localization does not confer a significant advantage to the embryo in this case.  In general, the transport of toxic molecules in the extracellular fluid will depend not only on diffusion, but also on fluid advection.  We quantify the fluid flow surrounding the embryo in Section III.  In Section IV we discuss the concentration boundary layer that forms when the toxin is advected along with the flow.  We revisit the first-passage problem in the case of strong advection in Section V.  A perturbative solution of the advection-diffusion equation in the regime of large P\'eclet number reveals that fluid flow enhances the effectiveness of the tip localization strategy.  In Section VI we discuss the effect of surface roughness on the first-passage probability.  We conclude in Section VII by highlighting connections between the model results and recent experiments on the development of sea urchin embryos.

\section{The case of pure diffusion}
Consider a spherical embryo of radius $R\sim 40 \,\mu m$.  In the absence of fluid flow, a toxin released from the tip of a microvilli will diffuse in the extracellular fluid.  The diffusion coefficient of the toxin in the extracellular fluid is $D \sim 10^{-5} \, cm^{2} s^{-1}$, characteristic of small molecules in water.  The goal is to determine the probability that a released toxin will be reabsorbed by the cell.  In this paper we consider the case of a perfect spherical absorber.  This approximation is not as severe as one might imagine, as the perfectly absorbing sphere is a relatively good approximation to the case of a patchy reactive surface \cite{berg1977physics}.  In what follows we do not treat the chemical kinetics associated with the absorption process.  In the model formulation, all molecules which reach the cell surface are absorbed.  This constitutes a {\it worst} case scenario for the cell.  As a result, the first-passage probability calculated will set an upper bound on the true absorption probability.  In addition, at the outset we will ignore reabsorption by the microvilli themselves, and only consider absorption by the spherical surface.  In this approximation, the only role of the microvilli is to release the toxin molecules at a distance $h$ above the surface of the cell.  In Section VI we will revisit this approximation and discuss the role of surface roughness on the absorption probability in more detail.   

The toxin concentration $C$ satisfies the diffusion equation
\begin{eqnarray}
\frac{\partial C}{\partial t} = D \nabla^{2}C.
\end{eqnarray}
Defining the dimensionless length $\xi=r/R$, concentration $c=R^{3}C$, and time $\tau=(Dt)/R^{2}$ yields
\begin{eqnarray}
\frac{\partial c}{\partial \tau} = \nabla_{\xi}^{2}c.
\end{eqnarray}
$\nabla_{\xi}^{2}c=\frac{1}{\xi^{2}}\frac{\partial}{\partial \xi}\left( \xi^{2}\frac{\partial c}{\partial \xi}\right) + \frac{1}{\xi^{2}\sin\theta} \frac{\partial}{\partial \theta} \left( \sin \theta \frac{\partial c}{\partial \theta}\right) + \frac{1}{\xi^{2}\sin^{2}\theta}\frac{\partial^{2}c}{\partial \phi^{2}}$ denotes the Laplacian with respect to the dimensionless radial variable $\xi$.  Considering the Laplace transform of the concentration $\tilde{c}=\int_{0}^{\infty} e^{-s\tau} c \,\mathrm{d}\tau$ gives the partial differential equation
\begin{eqnarray}
\nabla_{\xi}^{2}\tilde{c} - s \tilde{c} = -c(\tau=0) = - \delta^{3}(\vec{\xi} - \vec{\xi'}).  \label{difeq}
\end{eqnarray}
The initial condition corresponds to a point source at the microvilli tip, and reveals that the Laplace transform of the concentration is the Green's function for the modified Helmholtz operator.  A solution in spherical polar coordinates can be obtained by introducing the expansion
\begin{eqnarray}
\tilde{c} = \sum_{\ell=0}^{\infty}\sum_{m=-\ell}^{\ell} a_{\ell m}(\xi,\xi')Y_{\ell m}^{*}(\theta',\phi')Y_{\ell m}(\theta,\phi).  \label{expansion}
\end{eqnarray}
The resulting radial equation for $a_{\ell m}(\xi,\xi')$ is solved with the absorbing boundary condition at the cell surface $a_{\ell m}(\xi=1,\xi')=0$, and requiring the solution to be finite at infinity.  The solution can be expressed in terms of the spherical modified Bessel functions \cite{arfken1985mathematical} $i_{\ell}(x)=\sqrt{\frac{\pi}{2x}}I_{\ell+1/2}(x)$ and $k_{\ell}(x)=\sqrt{\frac{2}{\pi x}}K_{\ell+1/2}(x)$ as
\begin{eqnarray}
a_{\ell m}(\xi,\xi') = \gamma k_{\ell}(\gamma \xi_{>})\left[ i_{\ell}(\gamma \xi_{<}) - \frac{i_{\ell}(\gamma)}{k_{\ell}(\gamma)}k_{\ell}(\gamma \xi_{<})\right] \label{almsol}
\end{eqnarray}
where $\gamma^{2}=s$.  Here  $\xi_{<}$ ($\xi_{>}$) represents the smaller (larger) of $\xi$ and $\xi'$.  The first-passage probability is determined from the time integral of the diffusive current density impinging on the sphere surface, 
\begin{eqnarray}
\Pi_{D} = \int_{0}^{\infty} \mathrm{d}t \iint \vec{J} \cdot \vec{\mathrm{d}a}.
\end{eqnarray}
Evaluating $\vec{J}\cdot \vec{\mathrm{d}a}=D \left.\frac{\partial C}{\partial r}\right|_{r=R}R^{2}\, \sin \theta \,\mathrm{d} \theta \,\mathrm{d\phi}$ on the surface of the sphere, the first passage probability can be written simply in terms of the Laplace transform of the dimensionless concentration, 
\begin{eqnarray}
\Pi_{D} = \lim_{s\rightarrow 0} \int_{0}^{\pi} \sin \theta \, \mathrm{d}\theta \int_{0}^{2\pi} \mathrm{d}\phi \left. \frac{\partial \tilde{c}}{\partial \xi}\right|_{\xi=1}  = \frac{1}{\xi'}.  \label{difprob}
\end{eqnarray}
This remarkably simple and well known result \cite{redner2001guide} is illustrated in Fig. \ref{fpprobdiffonly}.  The details of the derivation are outlined in Appendix A.  The result indicates that, in the case of pure diffusion, tip localization is not a very effective strategy for reducing the chances that exported toxins get reabsorbed.  In the dimensionless coordinates, the tip of the microvilli is located at $\xi'=1+\frac{h}{R}$.  With microvilli of length $h\sim 2 \,\mu m$ and an embryo of radius $R\sim 40 \,\mu m$ the absorption probability is $\Pi_{D}=0.95$.  
Examining the structure of the microvilli solely through the lens of toxin export, if transport were diffusion limited, one might expect significantly longer microvilli than what is observed experimentally.  
\begin{figure}[h]
\includegraphics[width=3.5 in]{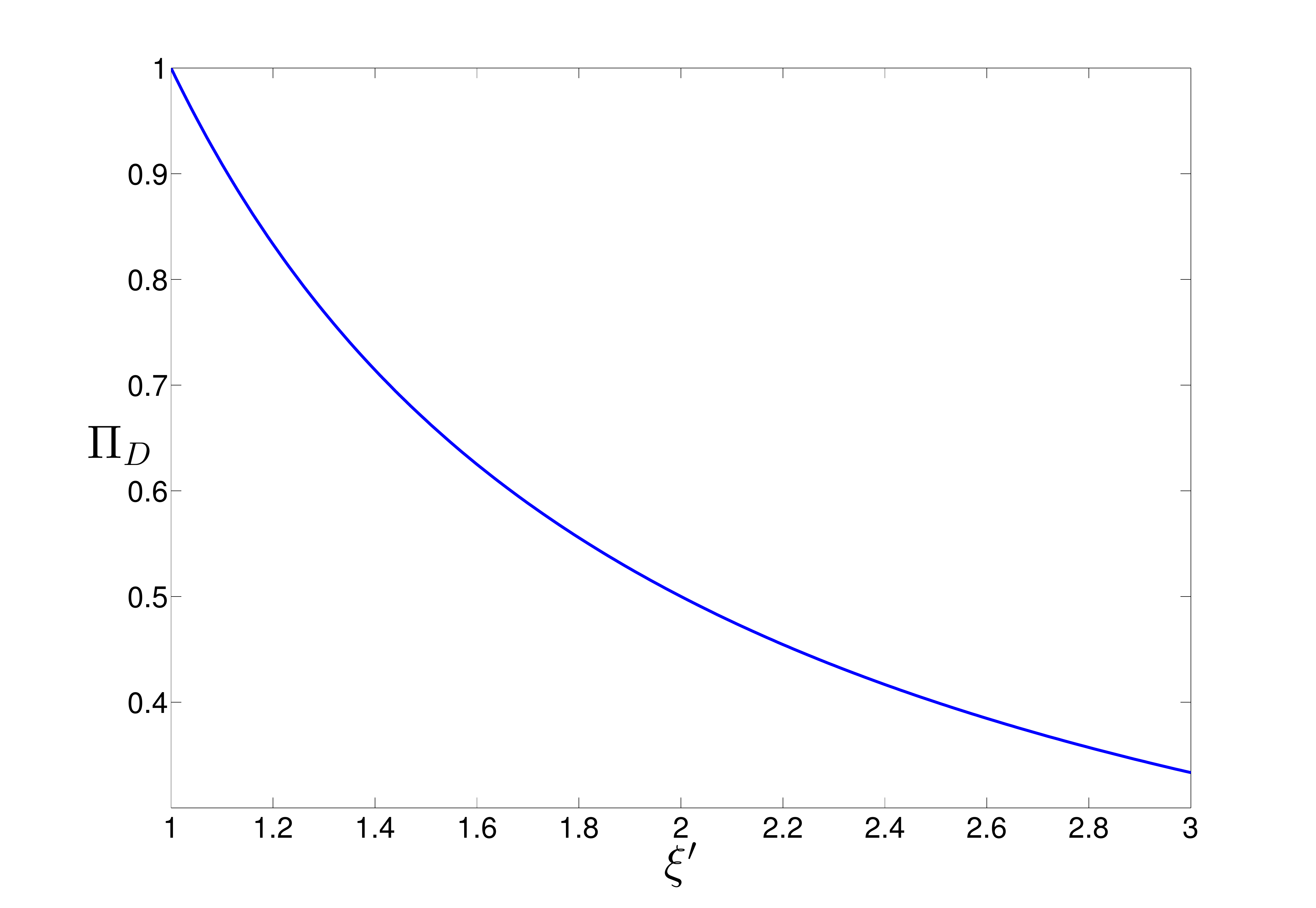}
\caption{\label{fpprobdiffonly}(Color online) The first-passage probability $\Pi_{D}$ as a function of the microvilli tip location $\xi'=1+\frac{h}{R}$ for the case of pure diffusion.  The microvilli have length $h$ and the embryo has radius $R$.  }
\end{figure}

\section{fluid flow}
In reality, the transport of toxins in the extra-cellular fluid is determined not only by diffusion, but also by advection.  The dimensionless P\'eclet number Pe characterizes the competition between advection and diffusion, 
\begin{align}
\text{Pe} = \frac{RU_{0}}{D}.
\end{align}
Here $U_{0}$ is a characteristic flow velocity, which will be discussed in more detail shortly.  We define the dimensionless fluid velocity as $\vec{u}=\frac{\vec{U}}{U_{0}}$.  For an incompressible fluid $\vec{\nabla}_{\xi} \cdot \vec{u}=0$ and the dimensionless toxin concentration satisfies
\begin{eqnarray}
\frac{\partial c}{\partial \tau} + \text{Pe} \, \vec{u}\cdot \vec{\nabla}_{\xi}c = \nabla_{\xi}^{2}c. \label{addif}
\end{eqnarray}
\\
An important property of the fluid flow is the Reynolds number
\begin{eqnarray}
\text{Re} = \frac{RU_{0}}{\nu} 
\end{eqnarray}
where $\nu \sim 10^{-6} m^{2}s^{-1}$ is the kinematic viscosity of ocean water.  To proceed we investigate the nature of the fluid flow in the vicinity of the embryo.  In particular, the wave swept rocky shore that is the habitat for the sea urchin is an environment where turbulent mixing takes place on the macroscale \cite{mead1995effects,denny1992biological}.  The question is how this turbulence manifests itself on the microscale of the embryo \cite{lazier1989turbulence,karp1996nutrient,berdalet2005effects}.  Kolmogorov's first similarity hypothesis states that the small scale fluid motion is universal and determined by two parameters, the kinematic viscostiy $\nu$ (units of [$m^{2}s^{-1}$]) and the turbulent kinetic energy dissipation rate $\varepsilon$ (units of [$m^{2}s^{-3}$]).  The unique length $\eta=(\nu^{3}/\varepsilon)^{1/4}$ and time $\tau_{\eta}=(\nu/\varepsilon)^{1/2}$ scales characterize the smallest dissipative eddies in the flow \cite{pope2000turbulent}.  In particular, the size of the smallest turbulent eddies is $\sim 2\pi\eta$ \cite{lazier1989turbulence}.  As a result, the smallest eddies are at least an order of magnitude larger than the embryo, and the local fluid environment of the embryo is one characterized by velocity gradients $\sim 1/\tau_{\eta}$.  

To calculate the first-passage probability we need to specify the specific form of the fluid velocity appearing in Eq. [\ref{addif}].  In what follows we will work with the model introduced earlier by Batchelor \cite{batchelor1979mass,batchelor1980mass}.  The model is applicable in the present case because Re $\ll$ 1, and we are considering the case of an isolated embryo.  For the calculation only the fluid velocity relative to the embryo matters.  This velocity is due in part to the motion of the embryo through the fluid as a result of an applied force and in part due to the ambient motion of the fluid which would be present even in the absence of the embryo.  The former takes into account gravity and includes the effect of bouyancy, since in general the density of the embryo will differ from that of the fluid.  One expects that in an otherwise quiescent fluid this density mismatch would lead a non-motile embryo to sink under the influence of gravity.  This behavior is observed experimentally in sea urchin embryos.  For example, the sinking velocity of \textit{Strongylocentrotus purpuratus} is $V \sim 0.4 \,mm\, s^{-1}$ \cite{mcdonald2004patterns}.  Interestingly this is comparable to the embryo's swimming velocity later in development.  The second contribution to the fluid velocity stems from the universal small scale motion of the fluid as a result of turbulent dissipation discussed above.  These two sources make independent contributions to the fluid velocity in the vicinity of the embryo.  Relative to the velocity of the embryo center, the fluid velocity $\vec{U}$ can be expressed as \cite{batchelor1980mass},
\begin{widetext}
\begin{eqnarray}
\vec{U} &= \vec{V} \cdot \left[ \left( \frac{3}{4\xi}+\frac{1}{4\xi^{3}}-1\right)\mathsf{I} + \left(\frac{3}{4\xi}-\frac{3}{4\xi^{3}} \right)\vec{\xi}\, \vec{\xi}\, \right] 
+R\,\mathsf{\Omega}\cdot \vec{\xi} + R\, \vec{\xi} \cdot \mathsf{E} \cdot \left[ \left(1-\frac{1}{\xi^{5}}\right) \mathsf{I} -\frac{5}{2} \frac{1}{\xi^{3}}\left(1-\frac{1}{\xi^{2}}\right)\vec{\xi}\, \vec{\xi}\,\right].
\end{eqnarray}
\end{widetext}
Here $\mathsf{I}$ is the unit tensor.  The first term accounts for the aforementioned sinking behavior due to gravity and the disturbance motion this generates in the flow.  As for the contribution from the ambient fluid motion (subscript $a$ for ambient),  the velocity gradient tensor $\vec{\nabla} \,\vec{U}_{a} = \mathsf{E} + \mathsf{\Omega}$ corresponds to the ambient fluid motion and has been decomposed into its symmetrical ($\mathsf{E}$) and antisymmetrical ($\mathsf{\Omega}$) parts.  The antisymmetric part $\mathsf{\Omega}_{ij} = - \frac{1}{2}\epsilon_{ijk}\omega_{k}$ represents the rigid body rotation of the embryo with angular velocity $\frac{1}{2}\vec{\omega}$ where $\vec{\omega} = \vec{\nabla} \times \vec{U}_{a}$ is the vorticity of the ambient flow \cite{batchelor1979mass}.  Here $\epsilon_{ijk}$ is the Levi-Civita symbol.  As discussed in \cite{batchelor1980mass}, in the low Reynolds number regime the embryo will rotate with the ambient fluid at all times.  In contrast, the embryo cannot follow the straining motion of the ambient fluid represented by the symmetric rate of strain tensor $\mathsf{E}$, which generates a disturbance motion in the flow.  

\begin{figure}[h]
\includegraphics[width=3.5 in]{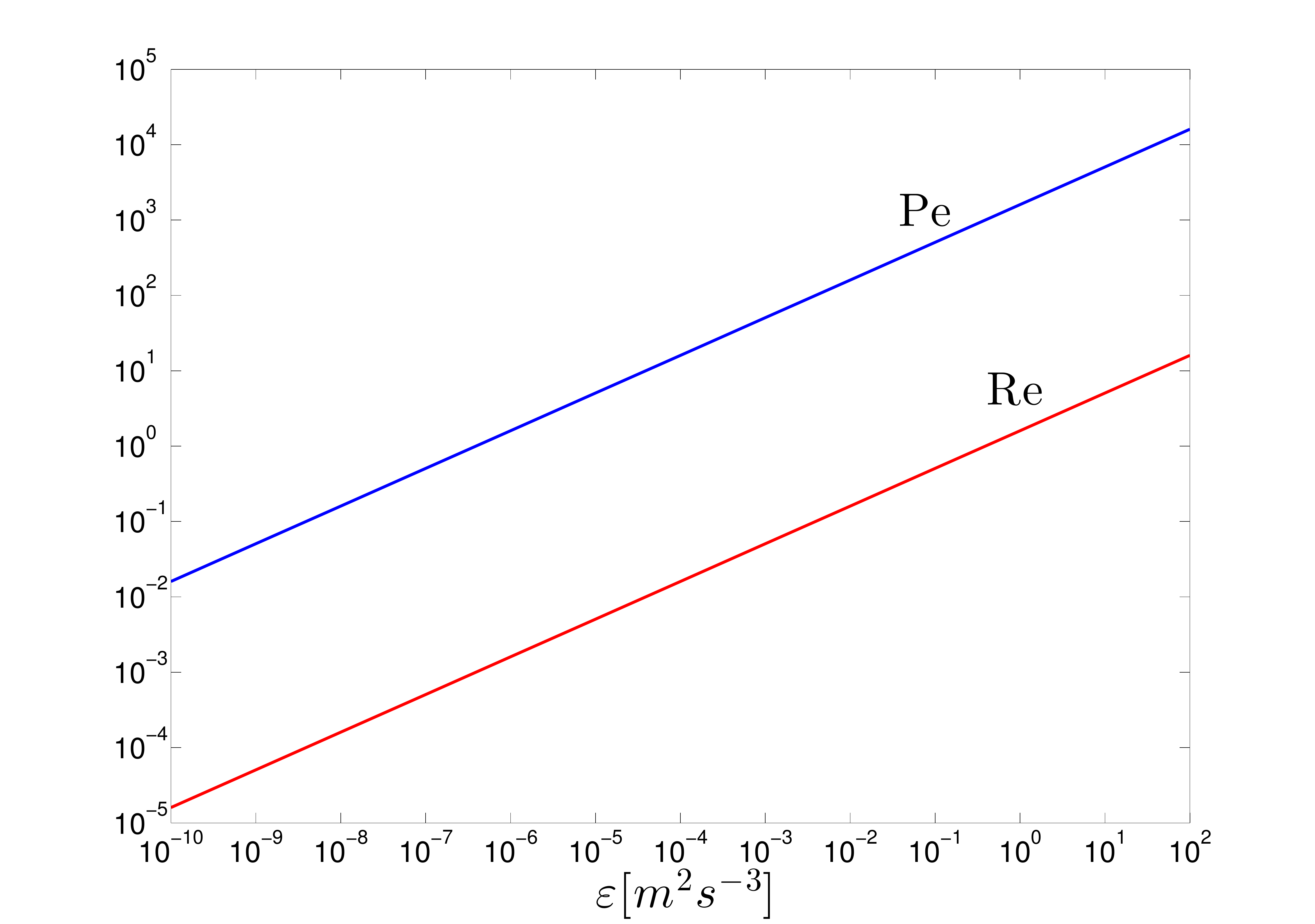}
\caption{\label{repe}(Color online) The dimensionless P\'eclet number $\text{Pe}$ (blue line) and Reynolds number $\text{Re}$ (red line) as a function of the turbulent kinetic energy dissipation rate $\varepsilon [m^{2}s^{-3}]$.  }
\end{figure}

This motivates defining the characteristic velocity $U_{0}=(R\omega)/2$ and hence the associated Reynolds number,
\begin{eqnarray}
\text{Re} = \frac{R^{2}\omega}{2\nu},
\end{eqnarray}
and P\'eclet number,
\begin{eqnarray}
\text{Pe} = \frac{R^{2}\omega}{2D}.
\end{eqnarray}
The microscale velocity gradient is related to the angular velocity as $1/\tau_{\eta}=\omega/2$.  We note that there is a great deal of variation, both spatial and temporal, in the value of $\varepsilon$ and hence $\omega$.  A characteristic value for the upper mixed layer of the ocean might be $\varepsilon \sim 10^{-6}\,m^{2}s^{-3}$ \cite{berdalet2005effects,lazier1989turbulence,karp1996nutrient}, whereas an embryo in a surge channel might be subject to instantaneous values a million times larger, $\varepsilon \sim 1 \,m^{2}s^{-3}$ \cite{denny1992biological,mead1995effects}.  
Using the value of the kinematic viscosity of ocean water, $\nu$, and an appropriate range of values for the kinetic energy dissipation rate, $\varepsilon$, one can see from Fig. \ref{repe} that the regime of interest is one where Re $\ll$ 1, but Pe $\gg$ 1.  Note that the condition Re $\ll$ 1 justifies the choice of a model in which the fluid velocity is obtained as a solution of Stokes equation.

In what follows we outline a program to calculate the first-passage probability perturbatively, making use of the fact that the quantity $\alpha=1/\text{Pe} \ll 1$.  
In spherical polar coordinates the resulting partial differential equation for the dimensionless concentration Eq. (\ref{addif}) is 
\begin{eqnarray}
\alpha \frac{\partial c}{\partial \tau} + u_{\xi}\frac{\partial c}{\partial \xi} +  \frac{u_{\theta}}{\xi} \frac{\partial c}{\partial \theta} + \frac{u_{\phi}}{\xi \sin \theta} \frac{\partial c}{\partial \phi}=\alpha \nabla_{\xi}^{2}c. \label{pde}
\end{eqnarray}
The dimensionless velocity components ($\vec{u}=\vec{U}/U_{0}$ and $\vec{v}=\vec{V}/U_{0}$) can be calculated as 
\begin{eqnarray}
u_{\xi}&=&(A+B)v_{\xi}+ (F - G)\mathsf{e}_{\xi\xi} \\
u_{\theta}&=& Av_{\theta} + F\mathsf{e}_{\xi\theta} \\
u_{\phi}&=& Av_{\phi} + \xi \sin \theta + F \mathsf{e}_{\xi\phi}
\end{eqnarray}
Here we have introduced the shorthand notation:
\begin{eqnarray}
A &=& \frac{3}{4\xi}+\frac{1}{4\xi^{3}}-1\\
B &=& \frac{3}{4\xi}\left(1-\frac{1}{\xi^{2}}\right)\\
F &=& \xi-\frac{1}{\xi^{4}}\\
G &=& \frac{5}{2\xi^{2}}\left(1-\frac{1}{\xi^{2}}\right)
\end{eqnarray}
In addition, we have introduced the dimensionless velocity gradient tensor $\vec{\nabla}_{\xi} \, \vec{u}_{a}=\mathsf{e} + \mathsf{\psi}$ with $\vec{u}_{a} = \vec{U}_{a}/U_{0}$, 
$\mathsf{e} = \frac{R}{U_{0}}\mathsf{E}$, and
$\mathsf{\psi}=\frac{R}{U_{0}}\mathsf{\Omega}$.
Note that the rotation of the embryo with the ambient fluid corresponds to $\mathsf{\Omega}_{\phi r}=\frac{1}{2}\omega \sin \theta$ and hence $\mathsf{\psi}_{\phi r} = \sin \theta$.  

The small quantity $\alpha$ multiplying the highest order spatial derivative in Eq. [\ref{pde}] is the hallmark of a boundary layer problem.  Physically this is an indication that the toxin concentration changes from its far field value to the value $c=0$ at the surface of the embryo ($\xi=1$) in a thin concentration boundary layer in the vicinity of the surface.  Within the concentration boundary layer, the dominant fluid motion is an azimuthal rotation, which corresponds to a solid body rotation of the embryo with the ambient fluid.  Superimposed on top of this rotation is a small fluctuation.  To proceed with the analysis we move to a reference frame rotating with the embryo, denoting the fluid velocity components in this frame by $\overset{\ast}{u}_{\gamma}$ with $\gamma \in \{\xi,\theta,\phi\}$.  The velocity components in the rotating frame can be obtained by removing the term $\xi \sin \theta$ from $u_{\phi}$, and making the replacement $\phi \rightarrow \phi - \text{Pe} \,\tau$.  In the rotating frame, defining a Cartesian coordinate system $(x_{1},x_{2},x_{3})$ with the $x_{3}$ direction along the direction of the ambient vorticity, the velocity components are obtained from the following relations:
\begin{widetext}
\begin{eqnarray}
\overset{\ast}{v}_{\xi} &=& v_{1} \sin \theta \cos(\phi- \text{Pe}\, \tau) + v_{2} \sin \theta \sin(\phi- \text{Pe}\, \tau)+ v_{3} \cos \theta \\
\overset{\ast}{v}_{\theta} &=& v_{1} \cos \theta \cos(\phi- \text{Pe}\, \tau) + v_{2} \cos \theta \sin(\phi- \text{Pe}\, \tau)- v_{3} \sin \theta \\
\overset{\ast}{v}_{\phi} &=& -v_{1} \sin(\phi- \text{Pe}\, \tau) + v_{2} \cos(\phi- \text{Pe}\, \tau) \\
\overset{\ast}{\mathsf{e}}_{\xi\xi} &=& \mathsf{e}_{11}\sin^{2}\theta \cos^{2}(\phi-\text{Pe}\, \tau) + \mathsf{e}_{22}\sin^{2}\theta \sin^{2}(\phi-\text{Pe}\, \tau) + \mathsf{e}_{33}\cos^{2}\theta + 
\mathsf{e}_{12}\sin^{2}\theta\sin(2(\phi-\text{Pe}\, \tau))+\nonumber \\
&& \mathsf{e}_{13}\sin(2\theta)\cos(\phi-\text{Pe}\, \tau)+ \mathsf{e}_{23}\sin(2\theta)\sin(\phi-\text{Pe}\, \tau) \\
\overset{\ast}{\mathsf{e}}_{\xi\theta} &=& \cos(2\theta) \left[ \mathsf{e}_{13} \cos(\phi-\text{Pe}\, \tau)+\mathsf{e}_{23} \sin(\phi-\text{Pe}\, \tau)\right]+ \nonumber \\ 
&&\frac{1}{4}\sin(2\theta)\left[\mathsf{e}_{11}+\mathsf{e}_{22}-2\mathsf{e}_{33}+(\mathsf{e}_{11}-\mathsf{e}_{22})\cos(2(\phi-\text{Pe}\, \tau)) + 2\mathsf{e}_{12}\sin(2(\phi-\text{Pe}\, \tau))\right] \\
\overset{\ast}{\mathsf{e}}_{\xi\phi} &=& \cos \theta [\mathsf{e}_{23} \cos(\phi-\text{Pe}\, \tau) - \mathsf{e}_{13}\sin(\phi-\text{Pe}\, \tau) ] + \mathsf{e}_{12}\sin \theta \cos(2(\phi-\text{Pe}\, \tau)) + \frac{1}{2}(\mathsf{e}_{22}-\mathsf{e}_{11})\sin(2(\phi-\text{Pe}\, \tau))
\end{eqnarray}
\end{widetext}

In principle the quantities $v_{i}$, $\mathsf{e}_{ij}$ ($\{i,j\}\in \{1,2,3\}$), and Pe are functions of time, fluctuating over a timescale $\tau \sim 1/\text{Pe}$ corresponding to the eddy turnover.  Following Batchelor \cite{batchelor1980mass}, we calculate the average velocity field in the vicinity of the embryo, by averaging over a timescale $\tau_{\text{long}} \gg 1/\text{Pe}$ that is long compared to the fluctuation timescale.  
\begin{eqnarray}
\langle\overset{\ast}{u}_{\gamma}\rangle=\frac{1}{\tau_{\text{long}} } \int_{0}^{\tau_{\text{long}} } \overset{\ast}{u}_{\gamma} \, \mathrm{d} \tau
\end{eqnarray}
Assuming that $v_{i}$, $\mathsf{e}_{ij}$, and Pe are stationary random functions of $\tau$, the average of many terms is zero, like $v_{i} \cos(\phi - \text{Pe}\, \tau)$ and $\mathsf{e}_{ij} \sin(\phi - \text{Pe}\, \tau)$.  
The result for the averaged components is:
\begin{eqnarray}
\langle\overset{\ast}{v}_{\xi}\rangle &=&  \langle v_{3} \rangle \cos \theta \\
\langle \overset{\ast}{v}_{\theta}\rangle &=&- \langle v_{3} \rangle \sin \theta \\
\langle \overset{\ast}{v}_{\phi} \rangle &=& 0 \\
\langle \overset{\ast}{\mathsf{e}}_{\xi\xi}\rangle  &=& \langle \mathsf{e}_{33} \rangle\\
\langle \overset{\ast}{\mathsf{e}}_{\xi\theta} \rangle &=& -\frac{3}{4} \sin(2\theta) \langle \mathsf{e}_{33} \rangle \\
\langle \overset{\ast}{\mathsf{e}}_{\xi\phi}\rangle  &=& 0
\end{eqnarray}
Here we have invoked the statistical isotropy of the small-scale turbulence, and the imcompressibility of the ambient fluid, $\mathsf{e}_{11}+\mathsf{e}_{22}+\mathsf{e}_{33}=0$.  As discussed in \cite{batchelor1980mass}, $\langle v_{3} \rangle = 0$.  As a result, the time-averaged, dimensionless velocity field depends on a single parameter $\langle \mathsf{e}_{33} \rangle$, which for locally homogeneous and isotropic turbulence takes on the value $\langle \mathsf{e}_{33} \rangle \simeq 0.18$.  
\begin{eqnarray}
\langle \overset{\ast}{u}_{\xi}\rangle &=&(F - G) \langle \mathsf{e}_{33}\rangle \\
\langle \overset{\ast}{u}_{\theta}\rangle&=& -\frac{3}{4} F \sin(2\theta) \langle \mathsf{e}_{33}\rangle \\
\langle \overset{\ast}{u}_{\phi}\rangle &=& 0
\end{eqnarray}

\begin{figure}[h]
\includegraphics[width=3.5 in]{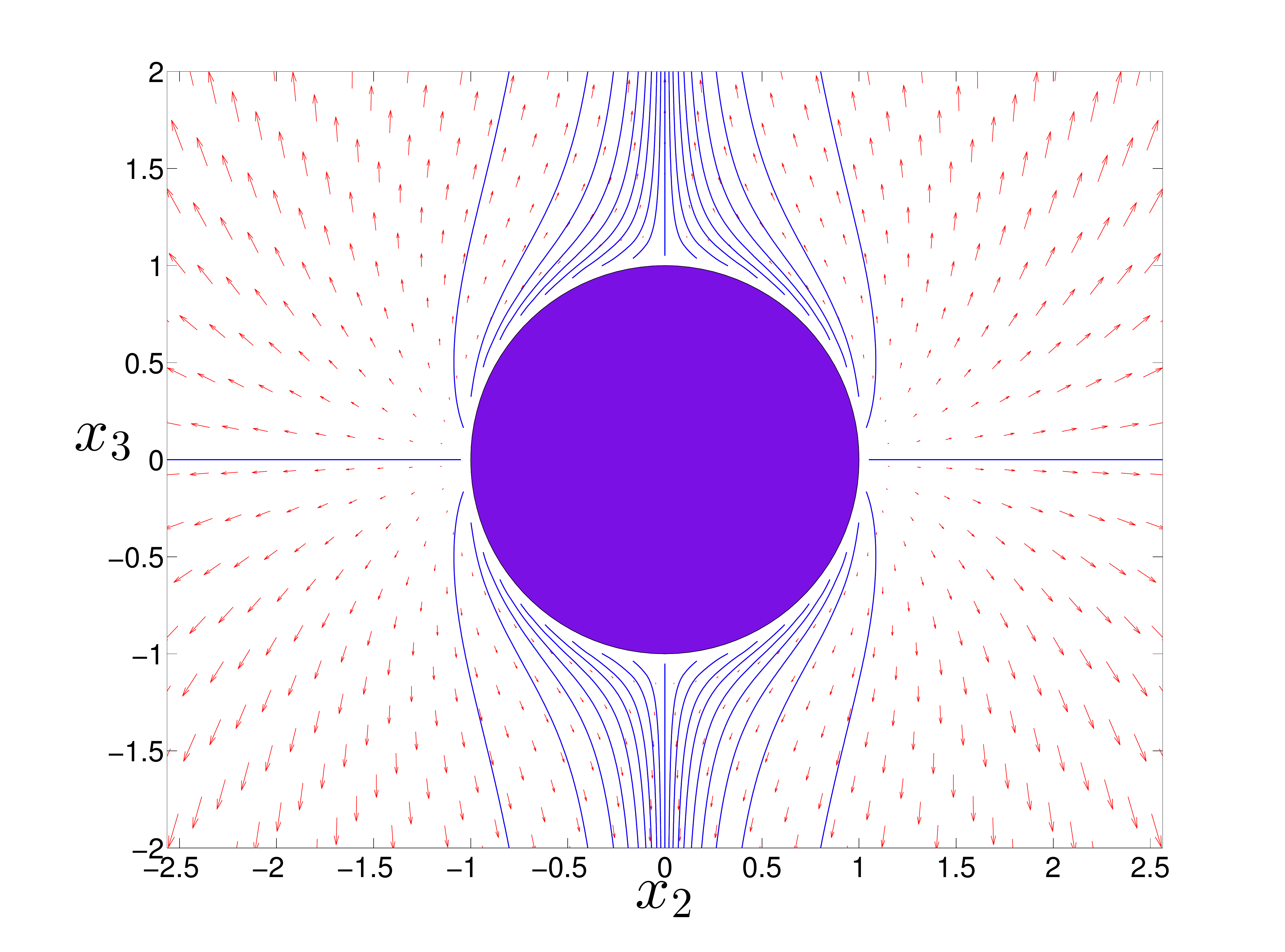}
\caption{\label{vecfield}(Color online) The time-averaged velocity field $\langle \overset{\ast}{\vec{u}}\rangle$ in the $(x_{2},x_{3})$ plane (red arrows).  Velocity streamlines starting at the tips of microvilli ($\xi'=1.05$) are shown as blue lines.  }
\end{figure}

The enhancement of mass transfer in the case of strong advection is now clear.  Within the concentration boundary layer, the average fluid flow consists of motion towards the north ($\theta < \pi/2$) or south ($\theta > \pi/2$) pole and a radial outflow (see Fig. \ref{vecfield}).  Toxin molecules released at the tips of microvilli will be advected away from the embryo, which will reduce their absorption probability.  

\section{concentration boundary layer}

\begin{figure}[h]
\includegraphics[width=3.5 in]{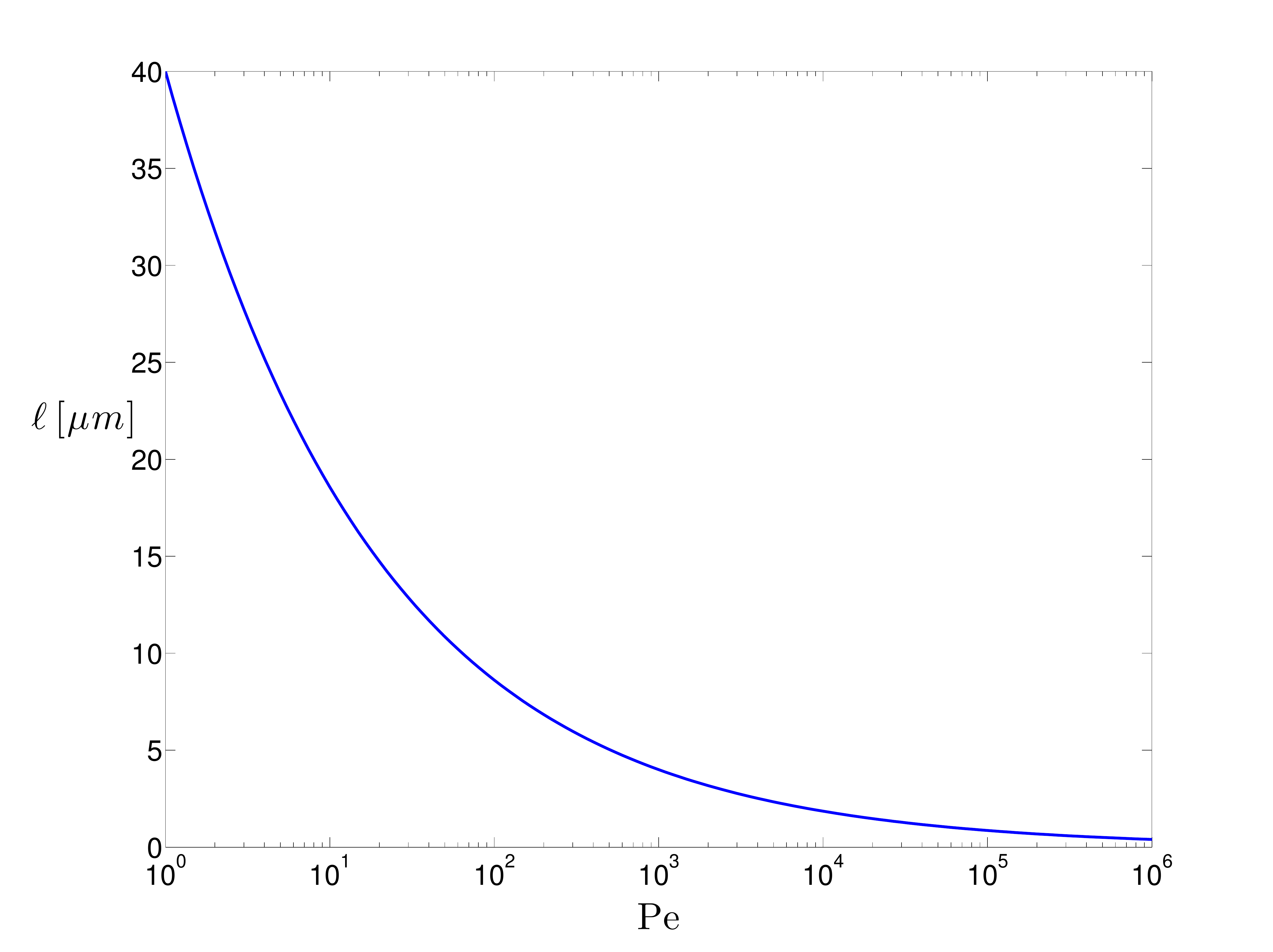}
\caption{\label{layerthick}(Color online) The concentration boundary layer thickness $\ell = R \,\text{Pe}^{-\frac{1}{3}}$ as a function of $\text{Pe}$.  At large $\text{Pe}$, the length of embryonic microvilli $h\approx \ell$.}
\end{figure}

With the time-averaged velocity field as input, the advection-diffusion equation for the toxin concentration is
\begin{eqnarray}
\alpha \frac{\partial c}{\partial \tau} +  \langle \overset{\ast}{u}_{\xi}\rangle\frac{\partial c}{\partial \xi} + \frac{\langle \overset{\ast}{u}_{\theta}\rangle}{\xi}\frac{\partial c}{\partial \theta} = \alpha \nabla_{\xi}^{2}c.
\end{eqnarray}
To investigate the quantiative implications of the boundary layer, we invoke the technique of dominant balance \cite{chen1996renormalization}.  Namely, we determine a rescaling of the radial variable $\xi=1+\alpha^{n}\rho$ which stretches out the boundary layer.  For the purposes of our first-passage calculation we find it useful to rescale the dimensionless time as $\tau = \alpha^{m}T$, but not the angular variables $\theta$ and $\phi$.  At this point the exponents $n$ and $m$ are unkown, but we are looking for a solution in which the lowest order governing equation for the concentration is independent of $\alpha$ and contains temporal, advective and diffusive terms.  The result of the rescaling is 
\begin{widetext}
\begin{eqnarray}
&\alpha^{2n-m}& \frac{\partial c}{\partial T} + \alpha^{n-1}\langle \overset{\ast}{u}_{\xi}\rangle \frac{\partial c}{\partial \rho} + \frac{\alpha^{2n-1}}{(1+\alpha^{n}\rho)} \langle \overset{\ast}{u}_{\theta} \rangle \frac{\partial c}{\partial \theta}   \nonumber \\ &=& \frac{\partial^{2}c}{\partial \rho^{2}}+ \frac{2\alpha^{n}}{(1+\alpha^{n}\rho)}\frac{\partial c}{\partial \rho}+\frac{\alpha^{2n}}{(1+\alpha^{n}\rho)^{2}}\frac{1}{\sin \theta}\frac{\partial}{\partial \theta}\left(\sin \theta \frac{\partial c}{\partial \theta}\right)+\frac{\alpha^{2n}}{(1+\alpha^{n}\rho)^{2}}\frac{1}{\sin^{2}\theta}\frac{\partial^{2}c}{\partial\phi^{2}}. \label{ordeq}
\end{eqnarray}
\end{widetext}

At this point it is important to remember (see Appendix B) that when expressed in terms of the radial variable $\rho$, the velocity components $\langle \overset{\ast}{u}_{\xi} \rangle\sim \mathcal{O}(\alpha^{2n})$ and $\langle \overset{\ast}{u}_{\theta}\rangle\sim \mathcal{O}(\alpha^{n})$ have nontrivial scaling with $\alpha$.  The correct choice of exponents for the rescaling is seen to be $n=1/3$ and $m=2/3$.  
In fact, the thickness of the concentration boundary layer, $\ell$, is determined by the exponent $n$ as (see Fig. \ref{layerthick})
\begin{eqnarray}
\ell = R \, \text{Pe}^{-\frac{1}{3}}.
\end{eqnarray}

We can now obtain a perturbative solution for the concentration in the form $c=\sum_{k=0}^{\infty}\alpha^{\frac{k}{3}}c_{k}$.  Inserting this expansion into Eq. [\ref{ordeq}] and collecting terms of the same order in $\alpha^{\frac{1}{3}}$ one obtains a system of coupled equations for the $\{c_{k}\}$.  Defining $\mu=\cos \theta$ and the parameter $\beta=\frac{15}{2}\langle \mathsf{e}_{33}\rangle$, the equations governing $c_{0}$ and $c_{1}$ are:
\begin{widetext}
\begin{eqnarray}
\frac{\partial c_{0}}{\partial T} + \beta \rho^{2}\frac{\partial c_{0}}{\partial \rho} + \beta \rho \mu (1-\mu^{2})\frac{\partial c_{0}}{\partial \mu} -\frac{\partial^{2} c_{0}}{\partial \rho^{2}} &=&0  \label{c0eq}\\
 \frac{\partial c_{1}}{\partial T} + \beta \rho^{2}\frac{\partial c_{1}}{\partial \rho} + \beta \rho \mu (1-\mu^{2})\frac{\partial c_{1}}{\partial \mu} -\frac{\partial^{2} c_{1}}{\partial \rho^{2}} &=& \frac{8}{3}\beta \rho^{3} \frac{\partial c_{0}}{\partial \rho} + 3 \beta \rho^{2} \mu (1-\mu^{2})\frac{\partial c_{0}}{\partial \mu} + 2 \frac{\partial c_{0}}{\partial \rho}\label{c1eq}
\end{eqnarray}
\end{widetext}

The perturbation program consists in calculating $c_{0}$ from Eq. (\ref{c0eq}), and using the solution to solve the inhomogeneous equation for $c_{1}$, Eq. (\ref{c1eq}).  The solutions for $c_{1}$ and $c_{0}$ can then utilized to calculate $c_{2}$, etc.  Following the common practice in boundary layer problems \cite{magar2003nutrient}, we define similarity variables $\eta = \rho/g$ and $\chi=T/g^{2}$, where the positive function $g(\mu)$ captures the angular dependence of the boundary layer.  In terms of this similarity transformation the zeroth order equation becomes
\begin{widetext}
\begin{eqnarray}
\frac{\partial c_{0}}{\partial \chi} + \beta \eta^{2} \left( g^{3} - \mu (1-\mu^{2})g^{2}\frac{\mathrm{d}g}{\mathrm{d}\mu}\right) \frac{\partial c_{0}}{\partial \eta} - \frac{\partial^{2}c_{0}}{\partial \eta^{2}} = 0.
\end{eqnarray}
\end{widetext}
Provided there is a solution where the term in brackets is equal to a constant, 
\begin{eqnarray}
g^{3} - \mu (1-\mu^{2})g^{2}\frac{\mathrm{d}g}{\mathrm{d}\mu}=\Delta,
\end{eqnarray}
the governing equation becomes
\begin{eqnarray}
\frac{\partial c_{0}}{\partial \chi} + \beta \Delta \eta^{2}  \frac{\partial c_{0}}{\partial \eta} - \frac{\partial^{2}c_{0}}{\partial \eta^{2}} = 0.
\end{eqnarray}
Without loss of generality we make the choice $\Delta=1$.  The differential equation for $g(\mu)$ is easily solved, with $\Upsilon$ a constant of integration,
\begin{eqnarray}
g(\mu) = \left( 1 + \Upsilon \frac{\mu^{3}}{(1-\mu^{2})^{\frac{3}{2}}}\right)^{\frac{1}{3}}.
\end{eqnarray}
We require that $g(\mu)$ be bounded, except at the poles $\mu=\pm 1$ where the boundary layer scaling may break down.  As a result we make the choice $\Upsilon=1$ for $\mu \geq 0$ and $\Upsilon = -1$ for $\mu <0$.

\begin{figure}[h]
\includegraphics[width=3.5 in]{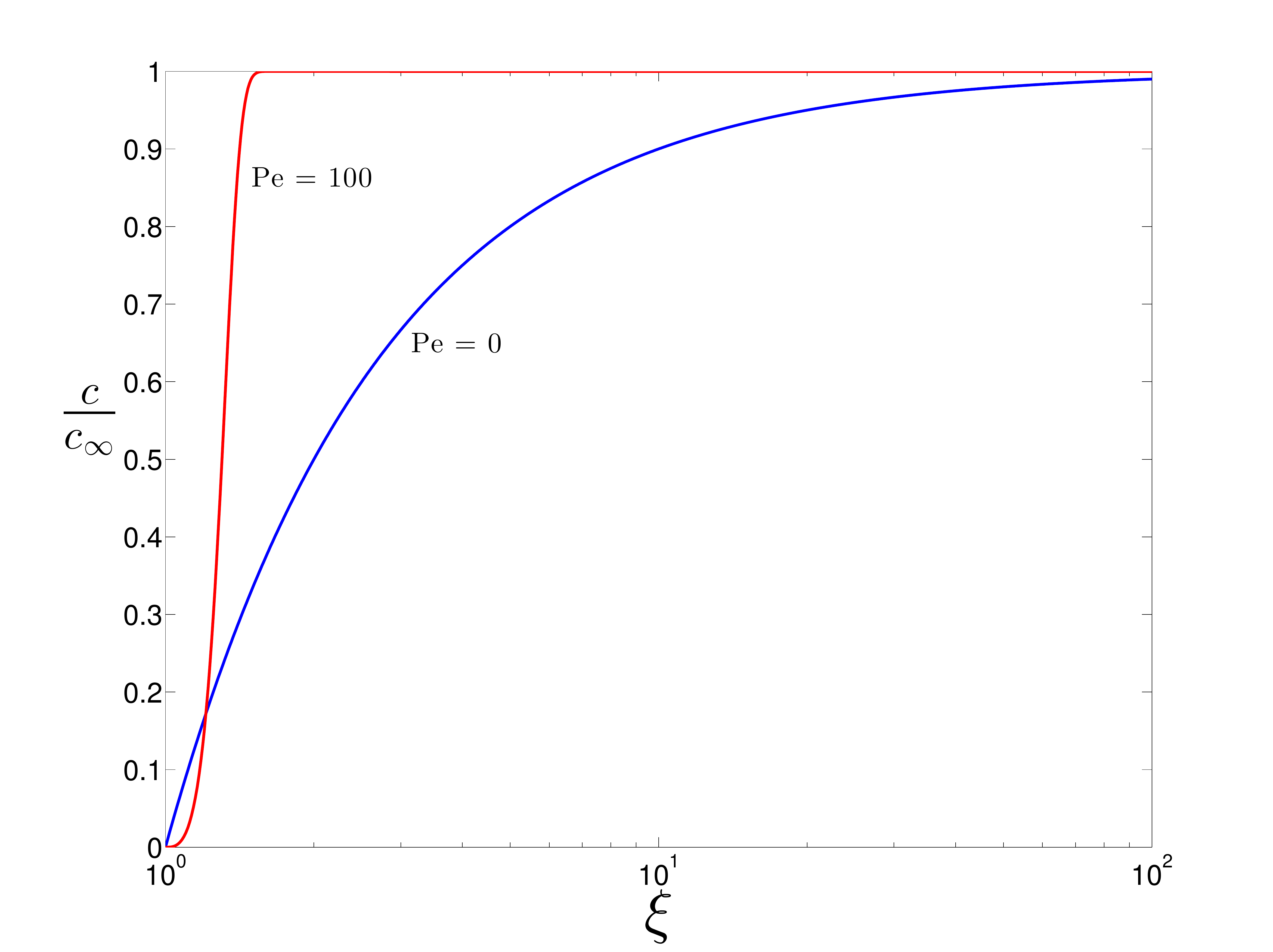}
\caption{\label{concprofile}(Color online) The steady-state concentration profile $c/c_{\infty}$, normalized by the far-field concentration $c_{\infty}$, as a function of the dimensionless radial variable $\xi$.  The case of pure diffusion ($\text{Pe}=0$), Eq. (\ref{difprof}), is shown as a blue line.  The concentration profile (along the line $\theta=\pi/4$) in the advection-dominated regime ($\text{Pe}=100$), Eq. (\ref{advecprof}), is shown as a red line.  }
\end{figure}

\begin{figure}[h]
\includegraphics[width=3.5 in]{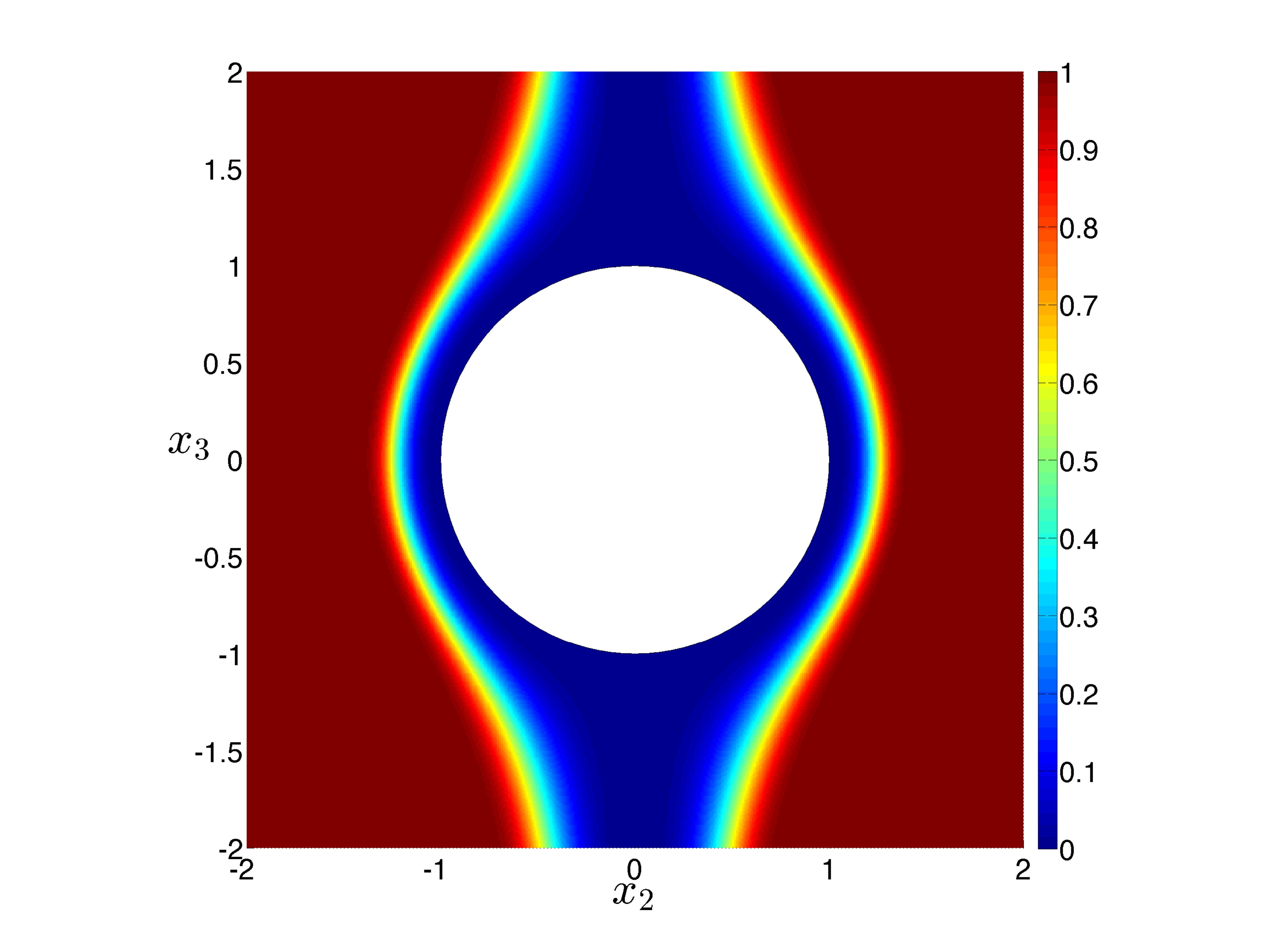}
\caption{\label{concprofile3}(Color online) Steady-state concentration contours of $c/c_{\infty}$ in the $(x_{2},x_{3})$ plane for the case of strong-advection ($\text{Pe}=100$).  Note the thickness of the concentration-boundary layer.  The concentration rapidly approaches its far field value ($c/c_{\infty}=1$) in a thin layer surrounding the embryo.  }
\end{figure}

Before tackling the first-passage problem, we highlight the physics of the concentration boundary layer by considering the steady-state solution ($\frac{\partial c_{0}}{\partial \chi}=0$) for the concentration profile in the presence of a perfectly absorbing sphere ($c_{0}=0$ at $\eta=0$) with toxin concentration $c_{\infty}$ far away from the sphere (see Figs. \ref{concprofile},\ref{concprofile3}).  The solution is readily obtained in terms of the incomplete Gamma function $\Gamma(a,z)$ as:
\begin{eqnarray}
c_{0} = c_{\infty}\left( 1- \frac{\Gamma(\frac{1}{3},\frac{\beta }{3}\eta^{3})}{\Gamma(\frac{1}{3})}\right) \label{advecprof}\\
\Gamma(a,z) = \int_{z}^{\infty} t^{a-1} e^{-t} \mathrm{d}t
\end{eqnarray}

\begin{figure}[h]
\includegraphics[width=3.5 in]{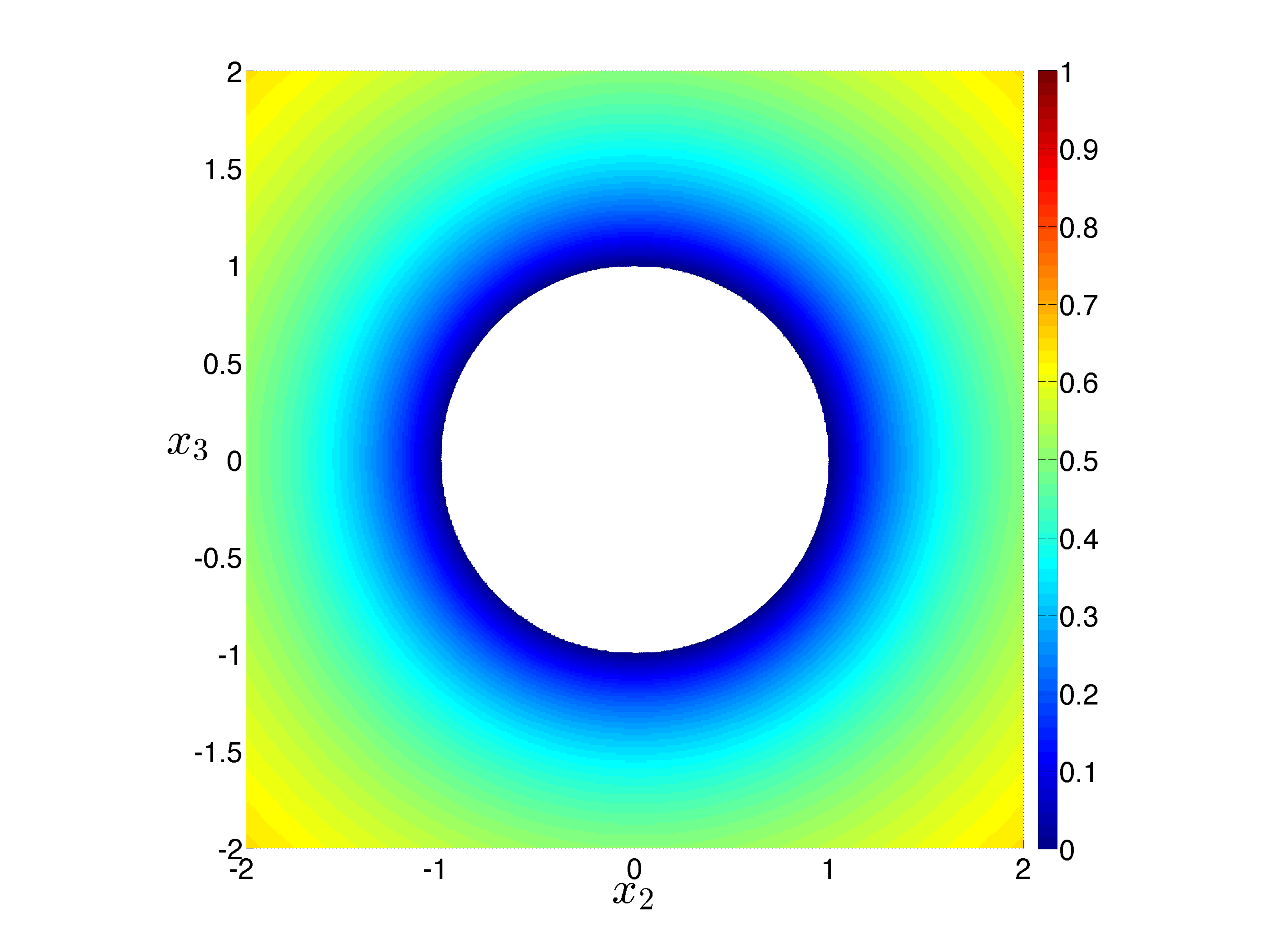}
\caption{\label{concprofile2}(Color online) Steady-state concentration contours of $c/c_{\infty}$ in the $(x_{2},x_{3})$ plane for the case of pure diffusion ($\text{Pe}=0$).  Note the thickness of the concentration-boundary layer.  At twice the embryo radius, the concentration has approached roughly half of its far field value ($c/c_{\infty}=1$).  }
\end{figure}

To quantify the mass transfer from the sphere in the case of strong advection, we calculate the zeroth order result for the dimensionless Sherwood number 
\begin{eqnarray}
\text{Sh}_{0} = \frac{1}{4\pi c_{\infty}} \int_{0}^{\pi} \sin \theta \,\mathrm{d}\theta \int_{0}^{2\pi} \mathrm{d} \phi \left. \frac{\partial c_{0}}{\partial \xi} \right|_{\xi=1}.  
\end{eqnarray}
Using the  above results we find:
\begin{eqnarray}
\text{Sh}_{0} &=& \left(\frac{3^{\frac{2}{3}}\beta^{\frac{1}{3}}\mathcal{I}}{2\,\Gamma(\frac{1}{3})}\right) \text{Pe}^{\frac{1}{3}}\approx 0.59 \, \text{Pe}^{\frac{1}{3}}\\
\mathcal{I} &=& \int_{-1}^{1} \frac{\mathrm{d}\mu}{g(\mu)} \approx 1.66
\end{eqnarray}
Defined in the same manner, $\text{Sh}_{D}=1$ for the case of pure diffusion ($\text{Pe}=0$), which can be readily obtained using the appropriate diffusive concentration profile $c = c_{\infty}\left( 1-\frac{1}{\xi}\right)$ for the given boundary conditions (see Fig. \ref{concprofile2}).  This highlights the advective enhancement of mass transfer away from the sphere $\sim \text{Pe}^{\frac{1}{3}}$ at $\text{Pe} \gg 1$, with the exponent $1/3$ coming from the boundary layer analysis.  

\section{the case of strong advection}

We now consider the first-passage problem for the case of strong advection.  For the purposes of the present calculation, we consider a spatial domain where all toxin molecules released at the tips of microvilli are eventually captured with probability one.  To do so, consider two perfectly absorbing surfaces, the first at the surface of the spherical embryo ($\eta = 0$), and a second at some prescribed distance ($\eta = \eta_{+}$).  We define the time-integrated concentration 
\begin{eqnarray}
\mathcal{C}_{0} = \int_{0}^{\infty} c_{0} \,\mathrm{d}\chi.
\end{eqnarray}
The equation governing $\mathcal{C}_{0}$ becomes 
\begin{eqnarray}
c_{0}(\chi = \infty) - c_{0}(\chi=0) + \beta \eta^{2}  \frac{\partial \mathcal{C}_{0}}{\partial \eta} - \frac{\partial^{2}\mathcal{C}_{0}}{\partial \eta^{2}} = 0.
\end{eqnarray}
Since all toxin molecules are absorbed with probability one, $c_{0}(\chi=\infty)=0$.  The initial condition corresponding to a point source at the microvilli tip is $c_{0}(\chi=0)=\delta^{3}(\vec{\xi}-\vec{\xi'})$.  By considering the sequence of variable transformations introduced earlier, $(\xi,\tau) \rightarrow (\rho, T) \rightarrow (\eta,\chi)$, and transforming the initial condition we arrive at the governing equation
\begin{widetext}
\begin{eqnarray}
 \frac{\partial^{2}\mathcal{C}_{0}}{\partial \eta^{2}} - \beta  \eta^{2}  \frac{\partial \mathcal{C}_{0}}{\partial \eta}= -\frac{\delta\left(\eta-\frac{g(\mu')}{g(\mu)}\eta'\right)\delta(\mu-\mu')\delta(\phi-\phi')}{\alpha^{\frac{1}{3}}g(\mu)\left(1+\alpha^{\frac{1}{3}}g(\mu)\eta\right)^{2}}. \label{C0govern}
\end{eqnarray}
The two independent solutions to the homogeneous equation (right hand side of Eq. ({\ref{C0govern}}) = 0) are a constant $\mathcal{C}_{0}^{(1)}= \kappa$, and the incomplete Gamma function $\mathcal{C}_{0}^{(2)}=\Gamma(\frac{1}{3},\frac{\beta}{3}\eta^{3})$.  The solution for $\mathcal{C}_{0}$ with absorbing boundary conditions can evidently be written in the form
\begin{eqnarray}
\mathcal{C}_{0} = Q \left(\Gamma\left(\frac{1}{3},\frac{\beta}{3}\eta_{<}^{3}\right)-\Gamma\left(\frac{1}{3}\right)\right) \left(  \Gamma\left(\frac{1}{3},\frac{\beta}{3}\eta_{>}^{3}\right) -\Gamma\left(\frac{1}{3},\frac{\beta}{3}\eta_{+}^{3}\right) \right).
\end{eqnarray}
Here $\eta_{<} \,(\eta_{>})$ is the smaller (larger) of $\eta$ and $\eta'$.  To determine the constant $Q$ we integrate both sides of the governing equation $\int_{-1}^{1}\mathrm{d}\mu \int_{0}^{2\pi}\mathrm{d}\phi\int_{\eta=\eta'-\epsilon}^{\eta=\eta'+\epsilon}\mathrm{d}\eta$ to determine the discontinuity in the first derivative of $\mathcal{C}_{0}$,
\begin{eqnarray}
4\pi \left. \frac{\partial\mathcal{C}_{0}}{\partial\eta}\right|_{\eta=\eta'-\epsilon}^{\eta=\eta'+\epsilon} = -\frac{1}{\alpha^{\frac{1}{3}}g(\mu')\left(1+\alpha^{\frac{1}{3}}g(\mu')\eta'\right)^{2}}.
\end{eqnarray}
A short calculation gives 
\begin{eqnarray}
Q = -\frac{e^{\frac{\beta}{3}(\eta')^{3}}}{4\pi 3^{\frac{2}{3}}(\beta\alpha)^{\frac{1}{3}}\left(\Gamma\left(\frac{1}{3}\right)-\Gamma\left(\frac{1}{3},\frac{\beta}{3}(\eta_{+})^{3}\right)\right)g(\mu')\left(1+\alpha^{\frac{1}{3}}g(\mu')\eta'\right)^{2}}.
\end{eqnarray}
\end{widetext}

\begin{figure}[h]
\includegraphics[width=3.5 in]{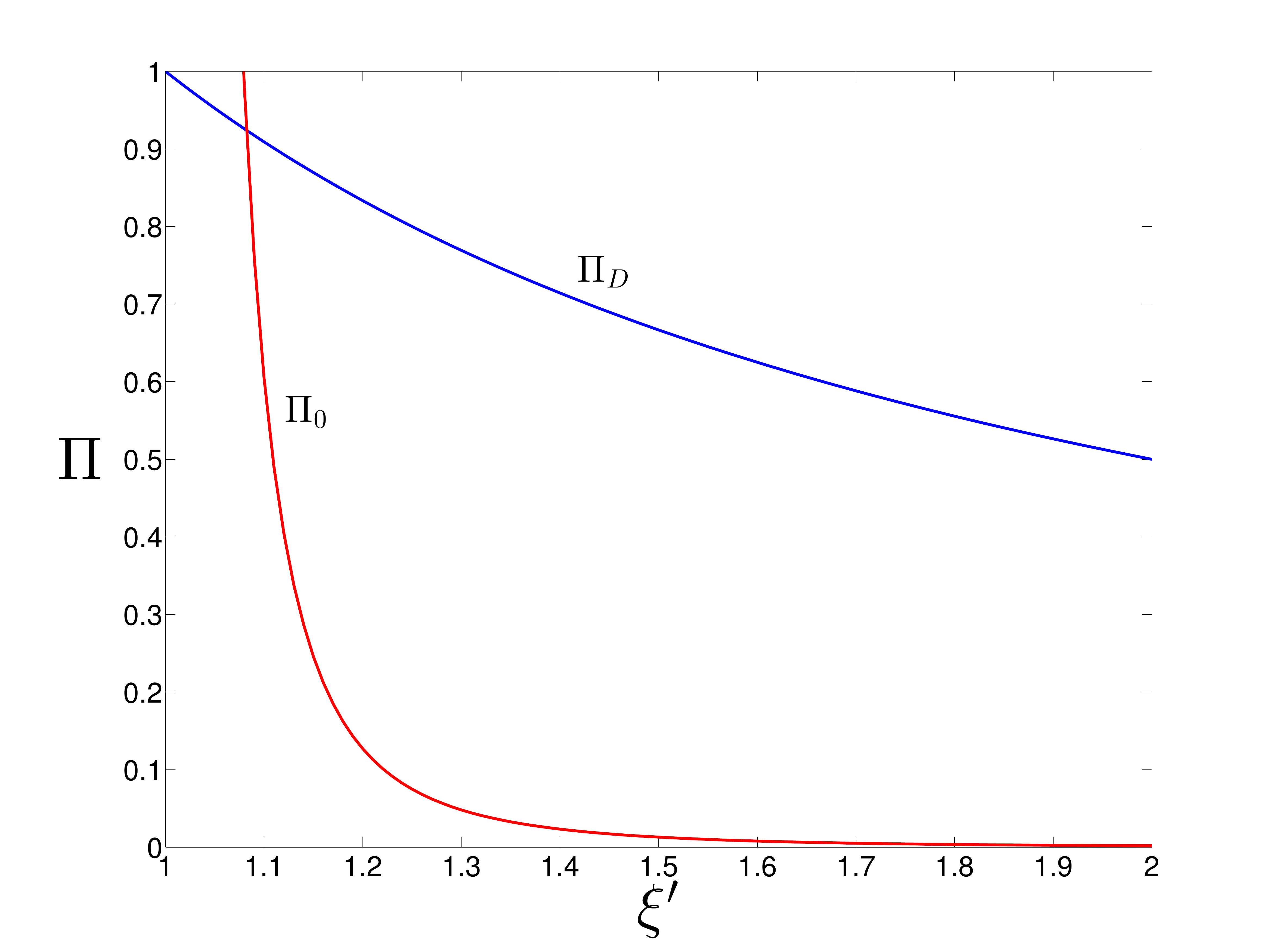}
\caption{\label{fpprob}(Color online) The first-passage probability $\Pi$ as a function of the microvilli tip location $\xi'$.  The result for the case of pure diffusion ($\text{Pe}=0$), $\Pi_{D}$, is shown as a blue line.  The zeroth order result for the case of strong-advection ($\text{Pe}=2062$), $\Pi_{0}$, is shown as a red line.  Note that in the advection-dominated case, the values of $\xi'$ over which there is a rapid decrease in absorption probability agree quite well with the length of embryonic microvilli.  Microvilli of height $h=2, 5, \text{and}\, 10 \,\mu m$ correspond to $\xi'=1.05, 1.125, \text{and}\, 1.25$, respectively.  }
\end{figure}

The first-passage probability is calculated from the concentration as 
\begin{eqnarray}
\Pi_{0} = \int_{0}^{\infty} \mathrm{d} \tau\int_{-1}^{1}\mathrm{d}\mu \int_{0}^{2\pi} \mathrm{d}\phi \left. \frac{\partial c_{0}}{\partial \xi}\right|_{\xi=1}.
\end{eqnarray}
Making the same sequence of variable transformations introduced earlier, the result can be written in terms of the time-integrated concentration $\mathcal{C}_{0}$ as 
\begin{eqnarray}
\Pi_{0} = \alpha^{\frac{1}{3}}\int_{-1}^{1}\mathrm{d}\mu \int_{0}^{2\pi}  \mathrm{d}\phi \left. \frac{\partial \mathcal{C}_{0}}{\partial \eta}\right|_{\eta=0} g(\mu).
\end{eqnarray}
The result of the angular integration gives:
\begin{widetext}
\begin{eqnarray}
\Pi_{0} &=& \frac{e^{\frac{\beta}{3}(\eta')^{3}}\Gamma\left(\frac{1}{3},\frac{\beta}{3}(\eta')^{3}\right)\mathcal{J}}{2\left(\Gamma\left(\frac{1}{3}\right)-\Gamma\left(\frac{1}{3},\frac{\beta}{3}(\eta_{+})^{3}\right)\right)g(\mu')\left(1+\alpha^{\frac{1}{3}}g(\mu')\eta'\right)^{2}} \\
\mathcal{J} &=& \int_{-1}^{1}\mathrm{d}\mu \,g(\mu) \approx 2.97
\end{eqnarray}
\end{widetext}
The result of the calculation can be greatly simplified by changing back to our original variables, and noting that $e^{z}\, \Gamma\left(\frac{1}{3},z\right)\approx z^{-\frac{2}{3}}+\mathcal{O}(z^{-\frac{4}{3}})$ for $z \gg 1$.  This approximation is justified in our case since $\alpha = 1/\text{Pe} \ll 1$ and therefore $z = \frac{\beta(\xi'-1)^{3}}{3\alpha g(\mu')^{3}} \gg 1$.  Taking the outer absorbing surface to infinity, $\eta_{+}\rightarrow \infty$, we arrive at the final result:
\begin{widetext}
\begin{eqnarray}
\Pi_{0} \approx \left(\frac{3^{\frac{2}{3}}\mathcal{J}}{2\Gamma(\frac{1}{3})\beta^{\frac{2}{3}}}\right)\frac{g(\mu')}{(\xi')^{2}(\xi'-1)^{2}}\,\text{Pe}^{-\frac{2}{3}} + \mathcal{O}(\text{Pe}^{-\frac{4}{3}}) \label{fp0}
\end{eqnarray}
\end{widetext}
The result can be interpreted simply as follows.  The first term in paranthesis is a dimensionless number of order unity, $\left(\frac{3^{\frac{2}{3}}\mathcal{J}}{2\Gamma(\frac{1}{3})\beta^{\frac{2}{3}}}\right)\approx0.94$, which depends on properties of the microscale velocity gradient ($\beta=\frac{15}{2}\langle \mathsf{e}_{33}\rangle$) and the angular dependence of the concentration boundary layer thickness (through $\mathcal{J}$).  The second term gives the dependence of the first-passage probability on the location ($\mu'$) and length ($\xi'$) of the microvillus that releases the toxin.  The last term gives the dependence of the first-passage probability on the P\'eclet number $\sim \text{Pe}^{-\frac{2}{3}}$.  Note the dramatic reduction (see Fig. \ref{fpprob}) of the first-passage probability as compared to the earlier case of diffusive transport ($\text{Pe}=0$), for which $\Pi_{D}=1/\xi'$.  

Comparing to the result in the case of pure diffusion, Eq. (\ref{difprob}), we see that in the advection dominated regime the first passage probability is reduced as compared to the purely diffusive first capture probability.  One can continue the perturbation program by calculating more of the $\{c_{k}\}$ and the leading order corrections to the first passage probability.  Provided that $\alpha^{\frac{1}{3}}=\text{Pe}^{-\frac{1}{3}}$ is small these corrections will not change the qualitative result of the zeroth order calculation.

The drastic reduction of uptake probability for microvilli lengths in quantitative agreement with experimental measurements of microvilli structure supports a functional significance to tip localization of toxin transporters.  When viewed through the lens of the toxin transport problem, one might say that the microvilli length has been evolutionarily selected to probe the thickness of the concentration boundary layer.  Toxin molecules released at the tips of microvilli will be advected away from the embryo, decreasing the probability that they will be reabsorbed and have to be exported again, which is energentically costly for the embryo.  Within the biological transporter literature, this sequence of export and subsequent reabsorption is refered to as futile cycling \cite{cole2013cost}.  

\begin{figure}[h]
\includegraphics[width=3.5 in]{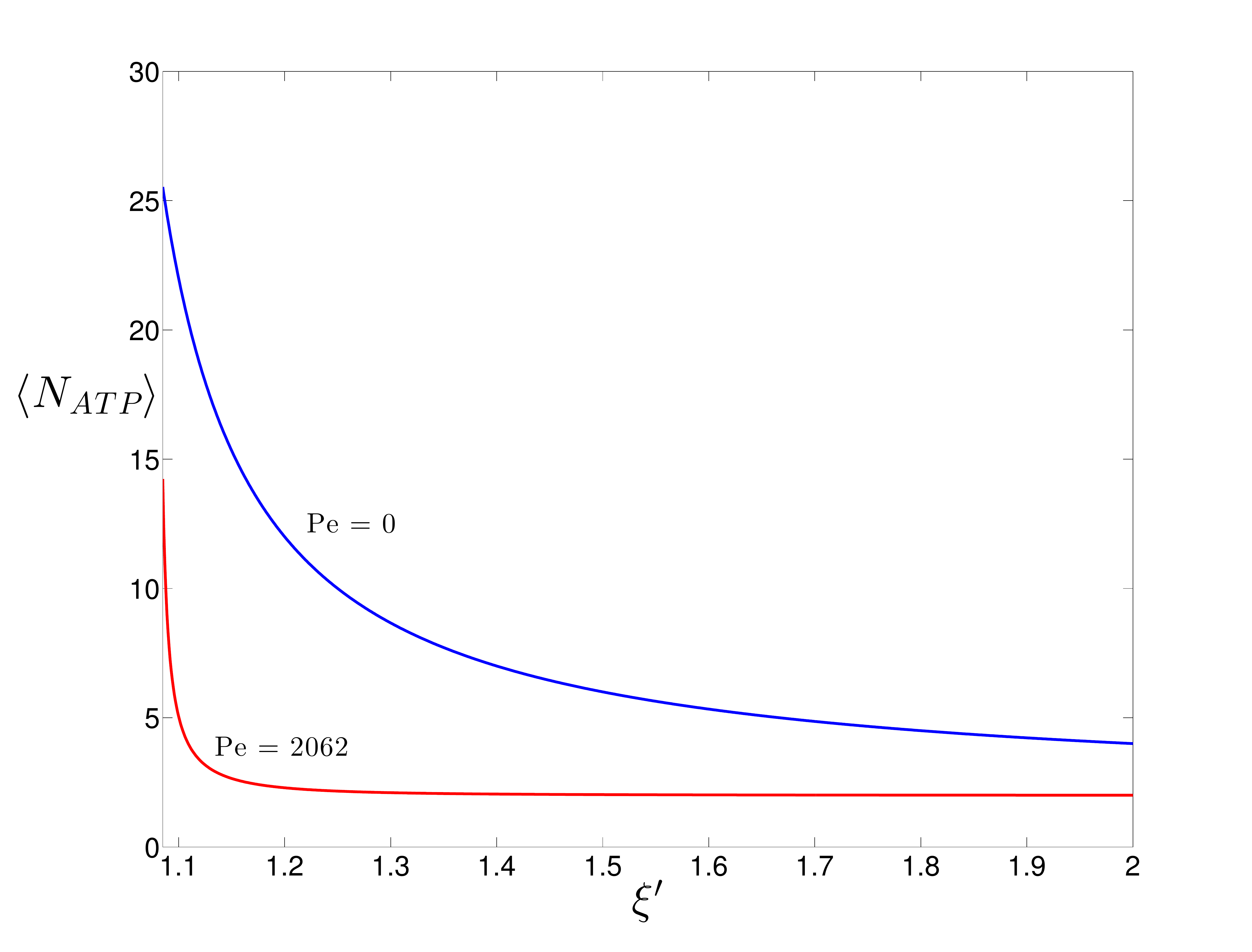}
\caption{\label{atpfig}(Color online) The average number of ATP molecules $\langle N_{ATP}\rangle$ required to efflux a single toxin molecule as a function of the microvilli tip location $\xi'$.  The result for the case of pure diffusion ($\text{Pe}=0$) is shown as a blue line.  The zeroth order result for the case of strong-advection ($\text{Pe}=2062$) is shown as a red line.  }
\end{figure}

Within the present first-passage formalism, we can quantify the cost associated with futile cycling of toxin molecules.  The cost to the embryo to efflux a single toxin molecule is two molecules of ATP.  As a result, the average number of ATP consumed to efflux a single toxin molecule is 
\begin{eqnarray}
\langle N_{ATP} \rangle = \sum_{n=1}^{\infty} 2n\Pi^{n-1}(1-\Pi) = \frac{2}{1-\Pi}.
\end{eqnarray}
As demonstrated in Fig. \ref{atpfig}, the effect of reducing the absorption probability is compounded when computing the cost of the transporter system for the embryo, with a significant reduction in the energy budget for the transporter system provided by the enhanced mass transprot at large Pe.  

\section{Surface roughness}
Thus far in our discussion, the role of the microvilli has been to simply displace the toxin above the surface of the embryo, where it is subsequently released into the extracellular fluid.  In our calculations of the first-passage probability, we have only considered absorption on the smooth spherical surface of the embryo.  In this approximation, the \textit{phantom} microvilli do not contribute to the surface area available for absorption, and do not modify the fluid flow in the vicinity of the embryo.  In this section we discuss how modifying these assumptions might affect the first-passage probabilities.  

To begin we collect some results about the microvillar architecture during sea urchin embryogenesis.  The microvilli are solitary, unbranched, cylindrical cell membrane protrusions.  There is substantial heterogeneity in the length of microvilli on the sea urchin embryo, with at least two populations of microvilli \cite{spiegel1989elongated,schroeder1978microvilli}.  The short microvilli (SMV) have a length of $h_{\text{SMV}}\simeq 2-3 \,\mu m$, comparable to the thickness of the hyaline layer that surrounds the embryo \cite{spiegel1990contractility,whalen2012actin}.  The elongated microvilli (EMV) are substantially longer, spanning the perivitelline space between the embryo surface and the fertilization envelope.  Their length depends on the width of the perivitelline space, in {\it Strongylocentrotus purpuratus} $h_{\text{EMV}}\simeq 35 \, \mu m$.  The radius $\varrho\simeq 0.1\, \mu m$ of the microvilli is the same for both populations (SMV and EMV).  According to studies on {\it Strongylocentrotus droebachiensis} there are $N\approx 3 \times 10^{5}$ microvilli covering the embryo \cite{schroeder1978microvilli}.  

The presence of microvilli increases the effective surface area of the embryo available for absorption and as a result should increase the first-passage probability.  The embryo's total surface area is 
\begin{eqnarray}
A_{\text{embryo}} = 4\pi R^{2} + N(2\pi \varrho h).
\end{eqnarray}
The first contribution is from the smooth spherical surface, and the second takes into account the cylindrical microvilli with average height $h$.  For an embryo with radius $R=40\,\mu m$, the smooth surface provides an area of $2.0\times10^{4}\, \mu m^{2}$.  With an average height of $h=2\,\mu m$, the microvilli provide an area of $3.8\times10^{5}\, \mu m^{2}$.  The result is that a rough embryo has a surface area at least 20 times as large as its smooth counterpart!  

To calculate the effect of surface roughness on the first-passage probability presents a significant challenge.  The technical problem is how the absorbing boundary condition can be applied on the rough surface.  An analytic approach to related problems has been developed based on ideas from multiple scattering theory \cite{foldy1945multiple}.  In principle the idea is to replace the exact boundary condition for the concentration $c$ on the rough surface (in our case the Dirichlet condition $c=0$ on the rough surface) by an effective boundary condition for the ensemble averaged concentration $\langle c \rangle_{\text{rough}}$ on the underlying smooth surface \cite{sarkar1995effective}.  The subscript ``rough" has been utilized so as not to confuse this averaging procedure with the temporal average utilized earlier in the paper for the computation of the fluid velocity.  The ensemble averaged concentration is defined as 
\begin{eqnarray}
\langle c \rangle_{\text{rough}} ( \vec{\xi}\,) = \frac{1}{N!}\int \mathrm{d} \mathcal{C}_{N} \,P(\mathcal{C}_{N}) \,c(\vec{\xi}\,|N).
\end{eqnarray} 
The notation $c(\vec{\xi}\,|N)$ emphasizes that the concentration depends not only on the position $\vec{\xi}$ but also on the configuration of the microvilli.  The averaging procedure is with respect to all possible arrangements of the microvilli on the smooth spherical surface.  Each arrangement of the microvilli is called a {\it configuration} denoted by $\mathcal{C}_{N}\equiv(\vec{Y}_{1},\vec{Y}_{2},...,\vec{Y}_{N})$.  Here $\vec{Y}_{i}$ denotes the position of the base of microvillus $i$ with respect to a curvilinear coordinate system on the smooth surface.  The normalization is defined by 
\begin{eqnarray}
N!=\int \mathrm{d}\mathcal{C}_{N}\, P(\mathcal{C}_{N}) = \int \mathrm{d}^{2}\vec{Y}_{1}\cdots \int \mathrm{d}^{2}\vec{Y}_{N} \,P(\mathcal{C}_{N}),
\end{eqnarray}
with a configuration appearing in the ensemble with probability $P(\mathcal{C}_{N})$.  The theory has been worked out in detail for the case of Laplace's equation \cite{sarkar1995effective}, $\nabla_{\xi}^{2}c=0$, which is the same as the steady-state diffusion equation.  The main result is an effective boundary condition for the ensemble averaged concentration, which, for a uniform spatial distribution of microvilli takes the form
\begin{eqnarray}
\left.\langle c \rangle_{\text{rough}} \right|_{\xi=1} = -\lambda \left. \frac{\partial \langle c \rangle_{\text{rough}}}{\partial \xi}\right|_{\xi=1}. \label{rough}
\end{eqnarray}
Note that the effect of surface roughness is to introduce a new lengthscale in the problem through the effective boundary condition.  The physical interpretation of $\lambda$ is a measure of the displacement of the $\langle c \rangle_{\text{rough}} = 0$ surface above the smooth surface.  In other words, if the Dirichlet boundary condition $c=0$ applies at the smooth surface $\xi=1$, the effect of surface roughness is to impose the condition $\langle c \rangle_{\text{rough}}=0$ at the surface $\xi=1+\lambda$.  Introducing the fraction of the smooth surface covered by the microvilli, $\varphi = \frac{N\pi \varrho^{2}}{4\pi R^{2}}$, the dimensionless length
\begin{eqnarray}
\lambda = (1+k)\varphi \frac{h}{R}.  
\end{eqnarray}
Here $k$ is a dimensionless number which in general depends on $\varphi$.  In the dilute limit, $\varphi \ll 1$, $k$ depends only on the shape of the microvilli.  

As an example to illustrate the potential effect of surface roughness, consider the solution of Laplace's equation $\nabla_{\xi}^{2} c=0$ for the dimensionless concentration $c$, with the Dirichlet boundary condition $c(\xi=1)=0$.  If the far field boundary condition is a constant concentration $c_{\infty}$, the solution is readily obtained as 
\begin{eqnarray}
c = c_{\infty}\left( 1- \frac{1}{\xi}\right). \label{difprof}
\end{eqnarray}
The dimensionless Sherwood number is calculated as 
\begin{eqnarray}
\text{Sh}_{D} = \frac{1}{4\pi c_{\infty}} \int_{0}^{\pi} \sin \theta \,\mathrm{d}\theta \int_{0}^{2\pi} \mathrm{d} \phi\left. \frac{\partial c}{\partial \xi} \right|_{\xi=1}=1.  
\end{eqnarray}
To determine the effect of surface roughness, consider the related problem for the ensemble averaged concentration $\langle c \rangle_{\text{rough}}$, with the Dirichlet boundary condition replaced by Eq. (\ref{rough}).  We calculate the concentration
\begin{eqnarray}
\langle c \rangle_{\text{rough}} = c_{\infty}\left(1 - \frac{1}{1-\lambda}\,\frac{1}{\xi}\right).  
\end{eqnarray}
The result for the Sherwood number is then
\begin{eqnarray}
\langle \text{Sh}_{D} \rangle_{\text{rough}} = \frac{1}{1-\lambda}.
\end{eqnarray}
The increase of toxin current density impining on the rough sphere should translate into an increase in the first-passage probability.  Unfortunately, a direct application of these results to the first-passage problem is somewhat problematic, since the effective boundary condition Eq. (\ref{rough}) is specific to the homogeneous Laplace equation.  For the first-passage application we would need results for Poisson's equation (for the case of pure diffusion), and the equation governing $\mathcal{C}_{0}$ (for the advection dominated regime).  An interesting avenue for future research is to extend the work of \cite{sarkar1995effective} to the present first-passage formalism.  

\begin{figure}[h]
\includegraphics[width=3.5 in]{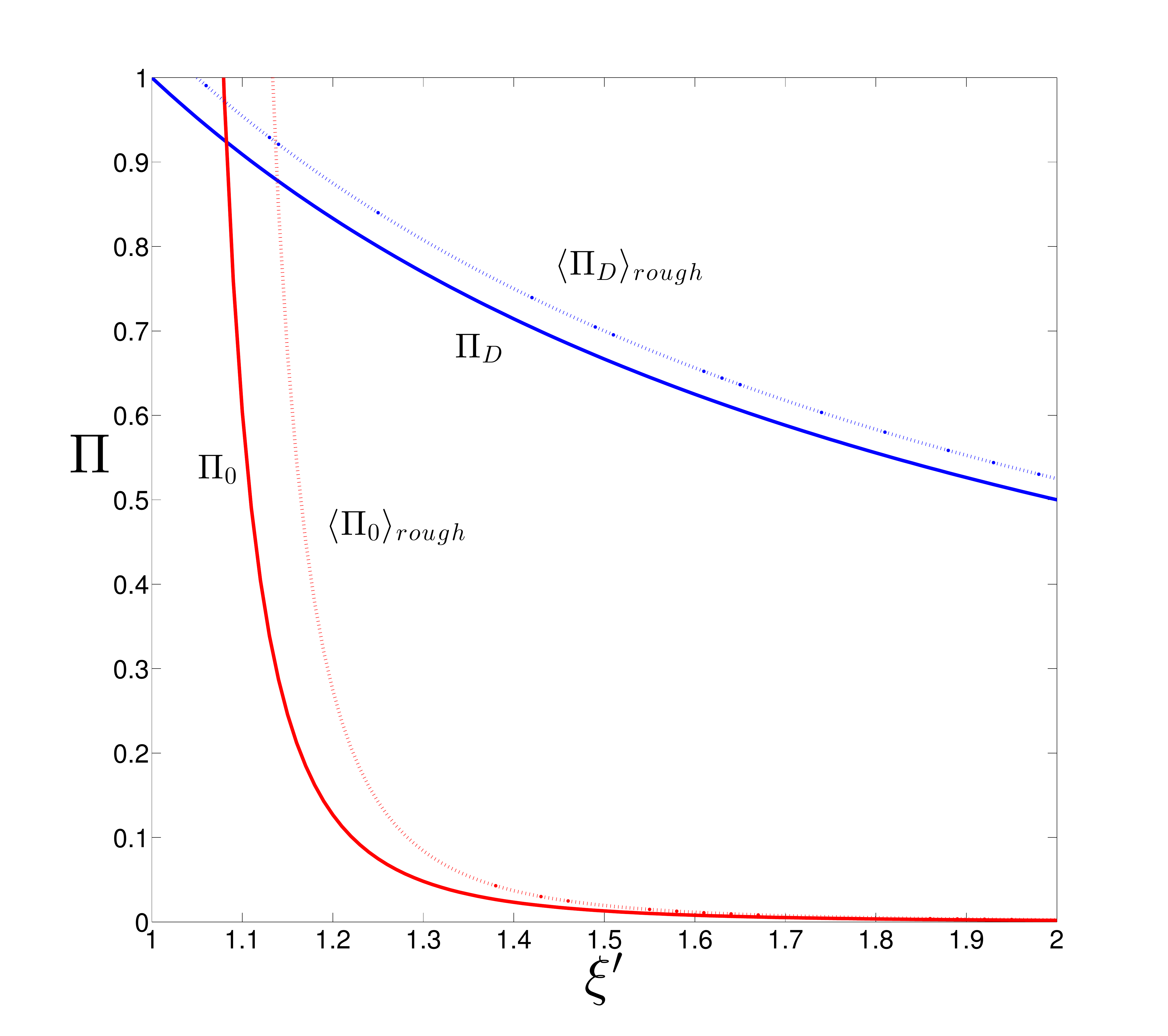}
\caption{\label{fprough}(Color online) The first passage probability $\Pi$ as a function of the microvilli tip location $\xi'$.  The effect of surface roughness is to increase the first-passage probability.  Heuristic estimates for the magnitude of the effect are provided by the dashed lines.   }
\end{figure}

 In what follows we consider a slightly more heuristic approach to capturing the effect of surface roughness.  Recall that the effective boundary condition can be interpreted as displacing the Dirichlet boundary condition above the smooth surface.  This suggests that we might be able to capture the effect of surface roughness by increasing the radius of the embryo and decreasing the length of the microvilli. 
\begin{eqnarray}
R_{\text{rough}} = R+\lambda R \\
h_{\text{rough}} = h-\lambda R
\end{eqnarray}
\\ Considering our earlier result for the diffusive first-passage probability, $\Pi_{D}=1/\xi'$, and recalling $\xi'=1+\frac{h}{R}$, after rescaling we find
\begin{eqnarray}
\langle \Pi_{D} \rangle_{\text{rough}} = \frac{1}{\xi_{\text{rough}}'} = \frac{(1+\lambda)}{\xi'} = (1+\lambda)\,\Pi_{D}.
\end{eqnarray}
This result is in agreement with  our calculation of the Sherwood number, which suggests enhancement by the factor $1/(1-\lambda)$, with deviations at $\mathcal{O}(\lambda^{2})$.  By performing the same rescaling (see Fig. \ref{fprough}), a naive extension to our result in the advection dominated regime suggests that 
\begin{widetext}
\begin{eqnarray}
\langle \Pi_{0} \rangle_{\text{rough}}\approx \left(\frac{3^{\frac{2}{3}}\mathcal{J}}{2\Gamma(\frac{1}{3})\beta^{\frac{2}{3}}}\right)\frac{g(\mu')}{\left(\frac{\xi'}{1+\lambda}\right)^{2}\left(\frac{\xi'}{1+\lambda}-1\right)^{2}}\,\text{Pe}^{-\frac{2}{3}} + \mathcal{O}(\text{Pe}^{-\frac{4}{3}}).
\end{eqnarray} 
\end{widetext}

To determine $\lambda$ we first calculate the surface fraction $\varphi \approx 0.47$.  This is not so small so as to safely rely on the dilute results for $k$, so as a first approximation 
we consider the numerical results derived at finite $\varphi$.  Note that the numerics are for the case of hemispherical microvilli \cite{sarkar1995effective}.  We find $1+k \approx 1.93$ and $\lambda\approx 0.05$.

We caution the reader that the discussion above is somehwat speculative, since the effecitve boundary condition Eq. (\ref{rough}) is specific to the homogeneous Laplace equation.  An interesting avenue for future research is to extend the work of \cite{sarkar1995effective} to the present first-passage formalism.  This would entail deriving an effective boundary condition similar to Eq. (\ref{rough}) for the inhomogeneous equations governing the concentration. Multiple scattering mehtods have also been applied to the problem of determining the disturbance in the flow field produced by surface roughness elements \cite{sarkar1996effective}.  The ensemble averaged flow field could then be utilized as input for the advection-diffusion equation to capture the effect of the microvilli on modifying the flow in the vicinity of the embryo surface.  This is a significant task for future research.  

\section{Conclusions}
In this paper we considered a spherical embryo of radius $R\sim 40\,\mu m$ in a flow-field with characteristic velocity $U_{0} \sim R/\tau_{\eta}$ as is typical for the smallest eddies in a turbulent macroscale flow.  The diffusion coefficient of the toxin in the extracellular fluid is $D \sim 10^{-5} \, cm^{2}\, s^{-1}$.  The dimensionless P\'eclet number which characterizes the competition between advection and diffusion is 
\begin{align}
\text{Pe} = \frac{RU_{0}}{D} \gg 1.
\end{align}
This means that relative to transport of the toxin in the extracellular fluid, advection is much more important than diffusion.  In this regime of large $\text{Pe}$, a concentration boundary layer forms near the embryo.  The boundary layer length scales as 
\begin{align}
\ell = R\, \text{Pe}^{-\frac{1}{3}}.
\end{align}
This gives a boundary layer of several microns in thickness.  Interestingly, this agrees quite well with the microvilli length, and would provide a physical reason for a distribution of transporters localized on the tips of the microvilli.  At the tips of the microvilli, the toxin concentration approaches the far-field value.  Toxins released at this height will be advected away from the embryo before having a chance to diffuse to the surface and be internalized.  The major result of the paper, Eq. (\ref{fp0}), is illustrated in Fig. \ref{fpprob}.  The argument is that the tip-localized transporter distribution and the microvilli architecture are evolutionarily adapted to probe the thickness of the concentration boundary layer.  The success and efficiency of the multi-drug transporters relies crucially on the presence of fluid flow in the open ocean environment of the sea urchin embryo.  Ignorant of the biochemical details of the transporter system, the physics governing mass transport at large P\'eclet number provides a compelling reason for the observed length of embryonic microvilli during sea urchin development.  

A number of simplifications have been made in the present paper.  For the purposes of building a tractable model system which does not obscure the underlying physics, many details of the sea urchin biology have been stripped away, including the presence of the hyaline layer surrounding the embryo and the fertilization envelope.  We have not considered how the microvilli will alter the fluid flow in the vicinity of the embryo.  Further work, in a computational fluid dynamics framework, could address these issues and incorporate a spatially varying toxin diffusivity.  Incorporating details of the chemical kinetics of the transporter system would pose a challenging problem of reaction, advection, and diffusion in a heterogeneous media.  

The major take home message from the paper on the relationship between the length scale of surface roughness elements and the mass transport problem is likely applicable beyond the scope of sea urchin development.  Villi are ubiquitous structures in biology \cite{kultti2006hyaluronan}, and similar ideas will carry over in other settings with a gradient in fluid velocity.  The design of a diverse variety of transport and mechanosensory systems may be guided by similar underlying principles, from toxin export by aquatic organisms residing in the benthic boundary layer \cite{bowden1978physical,mead1995effects}, to mechanotransduction by epithelial cells in the kidney \cite{guo2000hydrodynamic}.  

\section{acknowledgments}
This work was supported by a CASL Faculty Summer Research Grant from the University of Michigan-Dearborn.  Nicholas Licata acknowledges Sima Setayeshgar for insightful conversations about the sea urchin system.

\appendix
\section{The Case of Pure Diffusion (Pe = 0)}

This appendix outlines the solution for the first-passage probability in the purely diffusive case, where Pe = 0.  Using the completeness relation for the delta function in spherical polar coordinates
\begin{widetext}
\begin{eqnarray}
\delta^{3}(\vec{\xi}-\vec{\xi'}) = \frac{1}{\xi^{2}}\delta(\xi-\xi')\sum_{\ell=0}^{\infty}\sum_{m=-\ell}^{\ell}Y_{\ell m}^{*}(\theta',\phi')Y_{\ell m}(\theta,\phi)
\end{eqnarray}
and inserting the expansion Eq. (\ref{expansion}) into Eq. (\ref{difeq}) yields the radial equation
\begin{eqnarray}
\frac{\mathrm{d}^{2}a_{\ell m}}{\mathrm{d}\xi^{2}}+\frac{2}{\xi}\frac{\mathrm{d}a_{\ell m}}{\mathrm{d}\xi}-\frac{\ell(\ell+1)}{\xi^{2}}a_{\ell m}-\gamma^{2}a_{\ell m} = -\frac{1}{\xi^{2}}\delta(\xi-\xi'). \label{almeq}
\end{eqnarray}
Here we have defined $\gamma^{2}=s$.  Making the substitution $a_{\ell m}=\frac{b_{\ell m}}{(\gamma\xi)^{1/2}}$ the radial equation becomes
\begin{eqnarray}
\xi^{2}\frac{\mathrm{d}^{2}b_{\ell m}}{\mathrm{d}\xi^{2}}+\xi\frac{\mathrm{d}b_{\ell m}}{\mathrm{d}\xi}-\left[\left(\ell+\frac{1}{2}\right)^{2}+(\gamma\xi)^{2}\right]b_{\ell m} = -(\gamma\xi)^{\frac{1}{2}}\delta(\xi-\xi'). \label{blmeq}
\end{eqnarray}
As is evident from the form of the differential equation, the homogeneous solutions for the $b_{\ell m}$ are the modified Bessel functions of order $\ell+\frac{1}{2}$, denoted by $I_{\ell+\frac{1}{2}}(\gamma\xi)$ and $K_{\ell+\frac{1}{2}}(\gamma \xi)$.  The solution to Eq. (\ref{blmeq}) which is finite at infinity is
\begin{eqnarray}
b_{\ell m}(\xi,\xi') = K_{\ell+\frac{1}{2}}(\gamma\xi_{>})[A\,I_{\ell+\frac{1}{2}}(\gamma\xi_{<}) + B\, K_{\ell+\frac{1}{2}}(\gamma\xi_{<})].  
\end{eqnarray}
Here  $\xi_{<}$ ($\xi_{>}$) represents the smaller (larger) of $\xi$ and $\xi'$.  The absorbing boundary condition $b_{\ell m}=0$ at the surface of the embryo $\xi=1$ is satisfied by the choice $B=-A\frac{I_{\ell+\frac{1}{2}}(\gamma)}{K_{\ell+\frac{1}{2}}(\gamma)}$.  The remaining constant $A=(\frac{\gamma}{\xi'})^{1/2}$ is determined by integrating Eq. (\ref{blmeq}) from $\xi=\xi'-\epsilon$ to $\xi = \xi' + \epsilon$ and noting that the Wronskian of the modified Bessel functions is given by 
\begin{eqnarray}
I_{\ell+\frac{1}{2}}(x)\frac{\mathrm{d}K_{\ell+\frac{1}{2}}(x)}{\mathrm{d}x}-\frac{\mathrm{d}I_{\ell+\frac{1}{2}}(x)}{\mathrm{d}x}K_{\ell+\frac{1}{2}}(x)=-\frac{1}{x}.
\end{eqnarray}
Hence the solution for the $b_{\ell m}$ is
\begin{eqnarray}
b_{\ell m}(\xi,\xi') = \left(\frac{\gamma}{\xi'}\right)^{\frac{1}{2}}K_{\ell+\frac{1}{2}}(\gamma\xi_{>})\left(\,I_{\ell+\frac{1}{2}}(\gamma\xi_{<}) -\frac{I_{\ell+\frac{1}{2}}(\gamma)}{K_{\ell+\frac{1}{2}}(\gamma)}\, K_{\ell+\frac{1}{2}}(\gamma\xi_{<})\right).  
\end{eqnarray}
\end{widetext}
We define the spherical modified Bessel functions $i_{\ell}(x)=\sqrt{\frac{\pi}{2x}}I_{\ell+\frac{1}{2}}(x)$ and $k_{\ell}(x)=\sqrt{\frac{2}{\pi x}}K_{\ell+\frac{1}{2}}(x)$.  Note that the numerical factors in the definitions of $i_{\ell}(x)$ and $k_{\ell}(x)$ differ \cite{arfken1985mathematical}.  Making this substitution above and recalling the relation $a_{\ell m}=\frac{b_{\ell m}}{(\gamma\xi)^{1/2}}$, the solution to Eq. (\ref{almeq}) is Eq. (\ref{almsol}) from the main text,
\begin{eqnarray}
a_{\ell m}(\xi,\xi') = \gamma k_{\ell}(\gamma \xi_{>})\left[ i_{\ell}(\gamma \xi_{<}) - \frac{i_{\ell}(\gamma)}{k_{\ell}(\gamma)}k_{\ell}(\gamma \xi_{<})\right].
\end{eqnarray}
To calculate the first passage probability
\begin{eqnarray}
\Pi_{D} = \int_{0}^{\infty} \mathrm{d}t \iint \vec{J} \cdot \vec{\mathrm{d}a}
\end{eqnarray}
note that the current density $\vec{J}=-D\vec{\nabla} C$ and $\vec{\mathrm{d}a}=-\hat{r} \, R^{2} \sin \theta \,\mathrm{d}\theta \, \mathrm{d \phi}$.  Moving to the dimensionless variables introduced earlier the equation can be written as:
\begin{eqnarray}
\Pi_{D} &=& \lim_{s\rightarrow 0}\int_{0}^{\infty} e^{-s\tau} \jmath(\tau)\,\mathrm{d}\tau  = \lim_{s\rightarrow 0} \tilde{\jmath}(s)\\
\jmath(\tau) &=& \int_{0}^{\pi} \sin \theta \, \mathrm{d}\theta\ \int_{0}^{2\pi} \mathrm{d}\phi \left. \frac{\partial c}{\partial \xi}\right|_{\xi=1}
\end{eqnarray}
This establishes that the first-passage probability can be calculated from the Laplace transform of the current $\tilde{\jmath}(s)$ by taking the limit that $s\rightarrow 0$.  We calculate
\begin{widetext}
\begin{eqnarray}
\tilde{\jmath}(s) = \sum_{\ell=0}^{\infty}\sum_{m=-\ell}^{\ell} \gamma k_{\ell}(\gamma \xi')\left.\left(\frac{\partial i_{\ell}(\gamma\xi)}{\partial \xi}-\frac{i_{\ell}(\gamma)}{k_{\ell}(\gamma)}\frac{\partial k_{\ell}(\gamma\xi)}{\partial \xi}\right)\right|_{\xi=1}Y_{\ell m}^{*}(\theta',\phi')\int_{0}^{\pi} \sin \theta \,\mathrm{d}\theta \int_{0}^{2\pi} \mathrm{d}\phi \, Y_{\ell m}(\theta,\phi).
\end{eqnarray}
\end{widetext}
As a result of the angular integration $\int_{0}^{\pi} \sin \theta \,\mathrm{d}\theta \int_{0}^{2\pi} \mathrm{d}\phi \, Y_{\ell m}(\theta,\phi) = \sqrt{4\pi}\delta_{\ell,0} \delta_{m,0}$ the only nonzero term has $\ell=m=0$.  Using the fact that $i_{0}(x)=\sinh(x)/x$ and $k_{0}(x)=e^{-x}/x$ a short calculation gives
\begin{eqnarray}
\tilde{\jmath}(s) =\frac{e^{-\gamma(\xi'-1)}}{\xi'}.  
\end{eqnarray}
Recalling that $\gamma^{2}=s$ and taking the limit that $s\rightarrow 0$ of the above expression yields the final result quoted in the main text,
\begin{eqnarray}
\Pi_{D}=\frac{1}{\xi'}.  
\end{eqnarray}
\section{The Case of Strong Advection (Pe $\gg$ 1)}

This appendix provides details necessary for the solution for the first-passage probability in the case of strong advection, where Pe $\gg$ 1.  Defining a spherical polar coordinate system with the $x_{3}$-axis along the direction of the ambient vorticity, the Cartesian components of the antisymmetric part of the velocity gradient tensor take the form $\Omega_{ij}=-\frac{1}{2}\epsilon_{ij3}\omega$.  The spherical polar components are calculated as 
\begin{widetext}
\begin{eqnarray}
\left( \begin{array}{ccc}
\Omega_{rr} & \Omega_{r\theta} & \Omega_{r\phi} \\
\Omega_{\theta r} & \Omega_{\theta\theta} & \Omega_{\theta\phi} \\
\Omega_{\phi r} & \Omega_{\phi\theta} & \Omega_{\phi\phi} \end{array} \right) &=&
\left( \begin{array}{ccc}
\sin \theta \cos \phi & \sin \theta \sin \phi & \cos \theta \\
\cos \theta \cos \phi & \cos \theta \sin \phi & -\sin \theta \\
-\sin \phi & \cos \phi & 0 \end{array} \right)
\left( \begin{array}{ccc}
0 & - \frac{\omega}{2}  & 0\\
 \frac{\omega}{2}  & 0 & 0 \\
0 &0 & 0 \end{array} \right)
\left( \begin{array}{ccc}
\sin \theta \cos \phi & \cos \theta \cos \phi & -\sin \phi \\
\sin \theta \sin \phi & \cos \theta \sin \phi & \cos \phi \\
\cos \theta & -\sin \theta & 0 \end{array} \right) \nonumber \\
&=& \frac{\omega}{2}
\left( \begin{array}{ccc}
0 & 0 & -\sin \theta \\
0 & 0 & -\cos \theta \\
\sin \theta & \cos \theta & 0 \end{array} \right).
\end{eqnarray}
\end{widetext}
The spherical polar components of the symmetric part of the velocity gradient tensor $\mathsf{E}$ can be calculated in a similar fashion from the Cartesian components, which satisfy  $\mathsf{E}_{ij}=\mathsf{E}_{ji}$.  
Moving to a frame of reference that is rotating with the embryo by making the replacement $\phi \rightarrow \phi -\text{Pe}\,\tau$, the leading contributions to the time-averaged velpocity components expressed in terms of the radial variable $\rho$ are:
\begin{widetext}
\begin{eqnarray}
\langle \overset{\ast}{u}_{\xi} \rangle&=& \left(\frac{15}{2}\rho^{2} \,\alpha^{\frac{2}{3}} -20 \rho^{3} \,\alpha + \mathcal{O}\left(\alpha^{\frac{4}{3}}\right) \right) \langle \mathsf{e}_{33}\rangle\\
\langle \overset{\ast}{u}_{\theta} \rangle&=&  \left(-\frac{15}{4}\rho \,\alpha^{\frac{1}{3}}+\frac{15}{2}\rho^{2}\,\alpha^{\frac{2}{3}} -15\rho^{3} \,\alpha + \mathcal{O}\left(\alpha^{\frac{4}{3}}\right) \right) \sin(2\theta)\langle \mathsf{e}_{33}\rangle\\
\langle \overset{\ast}{u}_{\phi}\rangle &=& 0
\end{eqnarray}
\end{widetext}

\bibliographystyle{apsrev}
\bibliography{acompat,seaurchinbib}

\newif\ifabfull\abfulltrue
\begin{thebibliography}{31}
\expandafter\ifx\csname natexlab\endcsname\relax\def\natexlab#1{#1}\fi
\expandafter\ifx\csname bibnamefont\endcsname\relax
  \def\bibnamefont#1{#1}\fi
\expandafter\ifx\csname bibfnamefont\endcsname\relax
  \def\bibfnamefont#1{#1}\fi
\expandafter\ifx\csname citenamefont\endcsname\relax
  \def\citenamefont#1{#1}\fi
\expandafter\ifx\csname url\endcsname\relax
  \def\url#1{\texttt{#1}}\fi
\expandafter\ifx\csname urlprefix\endcsname\relax\def\urlprefix{URL }\fi
\providecommand{\bibinfo}[2]{#2}
\providecommand{\eprint}[2][]{\url{#2}}

\bibitem[{\citenamefont{Berg and Purcell}(1977)}]{berg1977physics}
\bibinfo{author}{\bibfnamefont{H.~C.} \bibnamefont{Berg}} \bibnamefont{and}
  \bibinfo{author}{\bibfnamefont{E.~M.} \bibnamefont{Purcell}},
  \bibinfo{journal}{Biophysical journal} \textbf{\bibinfo{volume}{20}},
  \bibinfo{pages}{193} (\bibinfo{year}{1977}).

\bibitem[{\citenamefont{Berg}(1993)}]{berg1993random}
\bibinfo{author}{\bibfnamefont{H.~C.} \bibnamefont{Berg}},
  \emph{\bibinfo{title}{Random walks in biology}}
  (\bibinfo{publisher}{Princeton University Press}, \bibinfo{year}{1993}).

\bibitem[{\citenamefont{Goldstein}(2011)}]{goldstein2011evolution}
\bibinfo{author}{\bibfnamefont{R.~E.} \bibnamefont{Goldstein}}, in
  \emph{\bibinfo{booktitle}{Biological Physics}}
  (\bibinfo{publisher}{Springer}, \bibinfo{year}{2011}), pp.
  \bibinfo{pages}{123--139}.

\bibitem[{\citenamefont{Short et~al.}(2006)\citenamefont{Short, Solari,
  Ganguly, Powers, Kessler, and Goldstein}}]{short2006flows}
\bibinfo{author}{\bibfnamefont{M.~B.} \bibnamefont{Short}},
  \bibinfo{author}{\bibfnamefont{C.~A.} \bibnamefont{Solari}},
  \bibinfo{author}{\bibfnamefont{S.}~\bibnamefont{Ganguly}},
  \bibinfo{author}{\bibfnamefont{T.~R.} \bibnamefont{Powers}},
  \bibinfo{author}{\bibfnamefont{J.~O.} \bibnamefont{Kessler}},
  \bibnamefont{and} \bibinfo{author}{\bibfnamefont{R.~E.}
  \bibnamefont{Goldstein}}, \bibinfo{journal}{Proceedings of the National
  Academy of Sciences} \textbf{\bibinfo{volume}{103}}, \bibinfo{pages}{8315}
  (\bibinfo{year}{2006}).

\bibitem[{\citenamefont{Whalen et~al.}(2012)\citenamefont{Whalen, Reitzel, and
  Hamdoun}}]{whalen2012actin}
\bibinfo{author}{\bibfnamefont{K.}~\bibnamefont{Whalen}},
  \bibinfo{author}{\bibfnamefont{A.~M.} \bibnamefont{Reitzel}},
  \bibnamefont{and} \bibinfo{author}{\bibfnamefont{A.}~\bibnamefont{Hamdoun}},
  \bibinfo{journal}{Molecular Biology of the Cell}
  \textbf{\bibinfo{volume}{23}}, \bibinfo{pages}{3663} (\bibinfo{year}{2012}).

\bibitem[{\citenamefont{G{\"o}kirmak et~al.}(2012)\citenamefont{G{\"o}kirmak,
  Campanale, Shipp, Moy, Tao, and Hamdoun}}]{gokirmak2012localization}
\bibinfo{author}{\bibfnamefont{T.}~\bibnamefont{G{\"o}kirmak}},
  \bibinfo{author}{\bibfnamefont{J.~P.} \bibnamefont{Campanale}},
  \bibinfo{author}{\bibfnamefont{L.~E.} \bibnamefont{Shipp}},
  \bibinfo{author}{\bibfnamefont{G.~W.} \bibnamefont{Moy}},
  \bibinfo{author}{\bibfnamefont{H.}~\bibnamefont{Tao}}, \bibnamefont{and}
  \bibinfo{author}{\bibfnamefont{A.}~\bibnamefont{Hamdoun}},
  \bibinfo{journal}{Journal of Biological Chemistry}
  \textbf{\bibinfo{volume}{287}}, \bibinfo{pages}{43876}
  (\bibinfo{year}{2012}).

\bibitem[{\citenamefont{Lange and Gartzke}(2001)}]{lange2001microvillar}
\bibinfo{author}{\bibfnamefont{K.}~\bibnamefont{Lange}} \bibnamefont{and}
  \bibinfo{author}{\bibfnamefont{J.}~\bibnamefont{Gartzke}},
  \bibinfo{journal}{American Journal of Physiology-Cell Physiology}
  \textbf{\bibinfo{volume}{281}}, \bibinfo{pages}{C369} (\bibinfo{year}{2001}).

\bibitem[{\citenamefont{Lange}(2011)}]{lange2011fundamental}
\bibinfo{author}{\bibfnamefont{K.}~\bibnamefont{Lange}},
  \bibinfo{journal}{Journal of cellular physiology}
  \textbf{\bibinfo{volume}{226}}, \bibinfo{pages}{896} (\bibinfo{year}{2011}).

\bibitem[{\citenamefont{Redner}(2001)}]{redner2001guide}
\bibinfo{author}{\bibfnamefont{S.}~\bibnamefont{Redner}},
  \emph{\bibinfo{title}{A guide to first-passage processes}}
  (\bibinfo{publisher}{Cambridge University Press}, \bibinfo{year}{2001}).

\bibitem[{\citenamefont{Arfken et~al.}(1985)\citenamefont{Arfken, Weber, and
  Ruby}}]{arfken1985mathematical}
\bibinfo{author}{\bibfnamefont{G.~B.} \bibnamefont{Arfken}},
  \bibinfo{author}{\bibfnamefont{H.-J.} \bibnamefont{Weber}}, \bibnamefont{and}
  \bibinfo{author}{\bibfnamefont{L.}~\bibnamefont{Ruby}},
  \emph{\bibinfo{title}{Mathematical methods for physicists}},
  vol.~\bibinfo{volume}{6} (\bibinfo{publisher}{Academic press New York},
  \bibinfo{year}{1985}).

\bibitem[{\citenamefont{Mead and Denny}(1995)}]{mead1995effects}
\bibinfo{author}{\bibfnamefont{K.~S.} \bibnamefont{Mead}} \bibnamefont{and}
  \bibinfo{author}{\bibfnamefont{M.~W.} \bibnamefont{Denny}},
  \bibinfo{journal}{The Biological Bulletin} \textbf{\bibinfo{volume}{188}},
  \bibinfo{pages}{46} (\bibinfo{year}{1995}).

\bibitem[{\citenamefont{Denny et~al.}(1992)\citenamefont{Denny, Dairiki, and
  Distefano}}]{denny1992biological}
\bibinfo{author}{\bibfnamefont{M.}~\bibnamefont{Denny}},
  \bibinfo{author}{\bibfnamefont{J.}~\bibnamefont{Dairiki}}, \bibnamefont{and}
  \bibinfo{author}{\bibfnamefont{S.}~\bibnamefont{Distefano}},
  \bibinfo{journal}{The Biological Bulletin} \textbf{\bibinfo{volume}{183}},
  \bibinfo{pages}{220} (\bibinfo{year}{1992}).

\bibitem[{\citenamefont{Lazier and Mann}(1989)}]{lazier1989turbulence}
\bibinfo{author}{\bibfnamefont{J.}~\bibnamefont{Lazier}} \bibnamefont{and}
  \bibinfo{author}{\bibfnamefont{K.}~\bibnamefont{Mann}},
  \bibinfo{journal}{Deep Sea Research Part A. Oceanographic Research Papers}
  \textbf{\bibinfo{volume}{36}}, \bibinfo{pages}{1721} (\bibinfo{year}{1989}).

\bibitem[{\citenamefont{Karp-Boss et~al.}(1996)\citenamefont{Karp-Boss, Boss,
  Jumars et~al.}}]{karp1996nutrient}
\bibinfo{author}{\bibfnamefont{L.}~\bibnamefont{Karp-Boss}},
  \bibinfo{author}{\bibfnamefont{E.}~\bibnamefont{Boss}},
  \bibinfo{author}{\bibfnamefont{P.}~\bibnamefont{Jumars}},
  \bibnamefont{et~al.}, \bibinfo{journal}{Oceanography and Marine Biology}
  \textbf{\bibinfo{volume}{34}}, \bibinfo{pages}{71} (\bibinfo{year}{1996}).

\bibitem[{\citenamefont{Berdalet and Estrada}(2005)}]{berdalet2005effects}
\bibinfo{author}{\bibfnamefont{E.}~\bibnamefont{Berdalet}} \bibnamefont{and}
  \bibinfo{author}{\bibfnamefont{M.}~\bibnamefont{Estrada}},
  \bibinfo{journal}{a Subba Rao, DV (Ed.). Algal Cultures, Analogues of Blooms
  and Applications} \textbf{\bibinfo{volume}{2}}, \bibinfo{pages}{459}
  (\bibinfo{year}{2005}).

\bibitem[{\citenamefont{Pope}(2000)}]{pope2000turbulent}
\bibinfo{author}{\bibfnamefont{S.~B.} \bibnamefont{Pope}},
  \emph{\bibinfo{title}{Turbulent flows}} (\bibinfo{publisher}{Cambridge
  university press}, \bibinfo{year}{2000}).

\bibitem[{\citenamefont{Batchelor}(1979)}]{batchelor1979mass}
\bibinfo{author}{\bibfnamefont{G.}~\bibnamefont{Batchelor}},
  \bibinfo{journal}{Journal of Fluid Mechanics} \textbf{\bibinfo{volume}{95}},
  \bibinfo{pages}{369} (\bibinfo{year}{1979}).

\bibitem[{\citenamefont{Batchelor}(1980)}]{batchelor1980mass}
\bibinfo{author}{\bibfnamefont{G.~K.} \bibnamefont{Batchelor}},
  \bibinfo{journal}{Journal of Fluid Mechanics} \textbf{\bibinfo{volume}{98}},
  \bibinfo{pages}{609} (\bibinfo{year}{1980}).

\bibitem[{\citenamefont{McDonald}(2004)}]{mcdonald2004patterns}
\bibinfo{author}{\bibfnamefont{K.}~\bibnamefont{McDonald}},
  \bibinfo{journal}{The Biological Bulletin} \textbf{\bibinfo{volume}{207}},
  \bibinfo{pages}{93} (\bibinfo{year}{2004}).

\bibitem[{\citenamefont{Chen et~al.}(1996)\citenamefont{Chen, Goldenfeld, and
  Oono}}]{chen1996renormalization}
\bibinfo{author}{\bibfnamefont{L.-Y.} \bibnamefont{Chen}},
  \bibinfo{author}{\bibfnamefont{N.}~\bibnamefont{Goldenfeld}},
  \bibnamefont{and} \bibinfo{author}{\bibfnamefont{Y.}~\bibnamefont{Oono}},
  \bibinfo{journal}{Physical Review E} \textbf{\bibinfo{volume}{54}},
  \bibinfo{pages}{376} (\bibinfo{year}{1996}).

\bibitem[{\citenamefont{Magar et~al.}(2003)\citenamefont{Magar, Goto, and
  Pedley}}]{magar2003nutrient}
\bibinfo{author}{\bibfnamefont{V.}~\bibnamefont{Magar}},
  \bibinfo{author}{\bibfnamefont{T.}~\bibnamefont{Goto}}, \bibnamefont{and}
  \bibinfo{author}{\bibfnamefont{T.}~\bibnamefont{Pedley}},
  \bibinfo{journal}{The Quarterly Journal of Mechanics and Applied Mathematics}
  \textbf{\bibinfo{volume}{56}}, \bibinfo{pages}{65} (\bibinfo{year}{2003}).

\bibitem[{\citenamefont{Cole et~al.}(2013)\citenamefont{Cole, Hamdoun, and
  Epel}}]{cole2013cost}
\bibinfo{author}{\bibfnamefont{B.~J.} \bibnamefont{Cole}},
  \bibinfo{author}{\bibfnamefont{A.}~\bibnamefont{Hamdoun}}, \bibnamefont{and}
  \bibinfo{author}{\bibfnamefont{D.}~\bibnamefont{Epel}}, \bibinfo{journal}{The
  Journal of experimental biology} \textbf{\bibinfo{volume}{216}},
  \bibinfo{pages}{3896} (\bibinfo{year}{2013}).

\bibitem[{\citenamefont{Spiegel et~al.}(1989)\citenamefont{Spiegel, Howard, and
  Spiegel}}]{spiegel1989elongated}
\bibinfo{author}{\bibfnamefont{E.}~\bibnamefont{Spiegel}},
  \bibinfo{author}{\bibfnamefont{L.}~\bibnamefont{Howard}}, \bibnamefont{and}
  \bibinfo{author}{\bibfnamefont{M.}~\bibnamefont{Spiegel}},
  \bibinfo{journal}{Roux's archives of developmental biology}
  \textbf{\bibinfo{volume}{198}}, \bibinfo{pages}{85} (\bibinfo{year}{1989}).

\bibitem[{\citenamefont{Schroeder}(1978)}]{schroeder1978microvilli}
\bibinfo{author}{\bibfnamefont{T.~E.} \bibnamefont{Schroeder}},
  \bibinfo{journal}{Developmental biology} \textbf{\bibinfo{volume}{64}},
  \bibinfo{pages}{342} (\bibinfo{year}{1978}).

\bibitem[{\citenamefont{Spiegel et~al.}(1990)\citenamefont{Spiegel, Howard, and
  Spiegel}}]{spiegel1990contractility}
\bibinfo{author}{\bibfnamefont{E.}~\bibnamefont{Spiegel}},
  \bibinfo{author}{\bibfnamefont{L.}~\bibnamefont{Howard}}, \bibnamefont{and}
  \bibinfo{author}{\bibfnamefont{M.}~\bibnamefont{Spiegel}},
  \bibinfo{journal}{Roux's archives of developmental biology}
  \textbf{\bibinfo{volume}{199}}, \bibinfo{pages}{228} (\bibinfo{year}{1990}).

\bibitem[{\citenamefont{Foldy}(1945)}]{foldy1945multiple}
\bibinfo{author}{\bibfnamefont{L.~L.} \bibnamefont{Foldy}},
  \bibinfo{journal}{Physical Review} \textbf{\bibinfo{volume}{67}},
  \bibinfo{pages}{107} (\bibinfo{year}{1945}).

\bibitem[{\citenamefont{Sarkar and Prosperetti}(1995)}]{sarkar1995effective}
\bibinfo{author}{\bibfnamefont{K.}~\bibnamefont{Sarkar}} \bibnamefont{and}
  \bibinfo{author}{\bibfnamefont{A.}~\bibnamefont{Prosperetti}},
  \bibinfo{journal}{Proceedings of the Royal Society of London. Series A:
  Mathematical and Physical Sciences} \textbf{\bibinfo{volume}{451}},
  \bibinfo{pages}{425} (\bibinfo{year}{1995}).

\bibitem[{\citenamefont{Sarkar and Prosperetti}(1996)}]{sarkar1996effective}
\bibinfo{author}{\bibfnamefont{K.}~\bibnamefont{Sarkar}} \bibnamefont{and}
  \bibinfo{author}{\bibfnamefont{A.}~\bibnamefont{Prosperetti}},
  \bibinfo{journal}{Journal of Fluid Mechanics} \textbf{\bibinfo{volume}{316}},
  \bibinfo{pages}{223} (\bibinfo{year}{1996}).

\bibitem[{\citenamefont{Kultti et~al.}(2006)\citenamefont{Kultti, Rilla,
  Tiihonen, Spicer, Tammi, and Tammi}}]{kultti2006hyaluronan}
\bibinfo{author}{\bibfnamefont{A.}~\bibnamefont{Kultti}},
  \bibinfo{author}{\bibfnamefont{K.}~\bibnamefont{Rilla}},
  \bibinfo{author}{\bibfnamefont{R.}~\bibnamefont{Tiihonen}},
  \bibinfo{author}{\bibfnamefont{A.~P.} \bibnamefont{Spicer}},
  \bibinfo{author}{\bibfnamefont{R.~H.} \bibnamefont{Tammi}}, \bibnamefont{and}
  \bibinfo{author}{\bibfnamefont{M.~I.} \bibnamefont{Tammi}},
  \bibinfo{journal}{Journal of Biological Chemistry}
  \textbf{\bibinfo{volume}{281}}, \bibinfo{pages}{15821}
  (\bibinfo{year}{2006}).

\bibitem[{\citenamefont{Bowden}(1978)}]{bowden1978physical}
\bibinfo{author}{\bibfnamefont{K.}~\bibnamefont{Bowden}},
  \bibinfo{journal}{Geophysical Surveys} \textbf{\bibinfo{volume}{3}},
  \bibinfo{pages}{255} (\bibinfo{year}{1978}).

\bibitem[{\citenamefont{Guo et~al.}(2000)\citenamefont{Guo, Weinstein, and
  Weinbaum}}]{guo2000hydrodynamic}
\bibinfo{author}{\bibfnamefont{P.}~\bibnamefont{Guo}},
  \bibinfo{author}{\bibfnamefont{A.}~\bibnamefont{Weinstein}},
  \bibnamefont{and} \bibinfo{author}{\bibfnamefont{S.}~\bibnamefont{Weinbaum}},
  \bibinfo{journal}{American Journal of Physiology-Renal Physiology}
  \textbf{\bibinfo{volume}{279}}, \bibinfo{pages}{F698} (\bibinfo{year}{2000}).

\end{thebibliography}
\end{document}